\documentclass[11pt]{amsart}
\usepackage[T1]{fontenc}
\usepackage{graphicx}
\usepackage[utf8]{inputenc}
\usepackage{url}
\usepackage{hyperref}
\input{xy}
\xyoption{all}
\xyoption{poly}
\usepackage[all]{xy}
\usepackage{float}
\usepackage{enumerate}
\usepackage[usenames]{color}
\usepackage{mathtools}
\usepackage{algorithm}
\usepackage{algpseudocode}
\usepackage{enumitem}

\def\D{{\mathcal D}}

\def\RR{{\mathbb R}}

\def\XX{{\mathbb X}}

\def\SWE{{\mathrm{SWE}}}
\def\VR{{\mathrm{VR}}}
\def\pers{{\mathrm{pers}}}
\def\Pers{{\mathrm{Pers}}}
\def\dgm{{\mathrm{dgm}}}

\usepackage{amsmath,amsthm,amsfonts,amssymb}

\theoremstyle{plain}

\newtheorem*{theorem*}{Theorem}

\theoremstyle{definition}

\theoremstyle{remark}

\title[Topological biomarkers for real-time detection of epileptic seizures]{Topological biomarkers for real-time detection of epileptic seizures}

\author[X. Fernández]{Ximena Fernández}
\address{Mathematical Institute, University of Oxford, and Department of Mathematical Sciences, Durham University, United Kingdom.}
\email{ximena.fernandez@maths.ox.ac.uk}
\author[D. Mateos]{Diego Mateos}
\address{Achucarro Basque Center For Neuroscience, Bizkaia, Spain and Universidad Aut\'onoma de Entre R\'ios (UADER) and Instituto de Matem\'atica Aplicada del Litoral (IMAL-CONICET-UNL), Argentina.}
\email{diego.mateos@achucarro.org}

\begin{document}

\begin{abstract}Real time seizure detection is a fundamental problem in computational neuroscience towards diagnosis and treatment's improvement  of epileptic disease. We propose a real-time computational method for  tracking and detection of epileptic seizures from raw neurophysiological recordings. Our mechanism is based on the topological analysis of the sliding-window embedding of the time series derived from simultaneously recorded channels. We extract topological biomarkers from the signals via the computation of the persistent homology of time-evolving topological spaces. Remarkably, the proposed biomarkers robustly captures the change in the brain dynamics during the ictal state. We apply our methods in different types of signals including scalp and intracranial electroencephalograms and magnetoencephalograms, in patients during interictal and ictal states, showing high accuracy in a range of clinical situations.\end{abstract}




\maketitle

\section{Introduction} 
Epilepsy is the second most common neurological disorder in the world after stroke, affecting an estimated fifty million people, according to the World Health Organization \cite{WHO22}. It is characterised by spontaneous seizures that occur when large regions of the brain exhibit synchronised neuronal activity and increased neuronal excitability \cite{shorvon2012oxford}. Understanding the mechanisms underlying the transition from a normal brain state to epileptic activity is crucial for both diagnosis and treatment. This is particularly important because about a third of epilepsy patients do not respond to conventional drug therapies and may require surgical intervention. Therefore, theoretical insights into these mechanisms are essential for the development of effective therapeutic strategies.
 
Diagnosis of epilepsy in hospitals typically relies on the expertise of neurophysiologists who analyze signals of brain activity obtained by electroencephalograms (EEG), intracranial EEG (iEEG), or magnetoencephalograms (MEG) \cite{P18}. While these methods serve as the gold standard, the interpretation of EEG, iEEG, or MEG recordings during the preictal, ictal, and interictal periods still depends largely on human judgment, which may be  influenced by various factors such as surgical sampling and training biases \cite{flanary2023reliability}. Despite decades of accumulated medical expertise, diagnosis by visual observation of recordings relies primarily on identifying changes in the signal waveform at a local level. However, this approach can be limited when dealing with a significant amount of information within the recording. Therefore, the integration of mathematical tools into signal analysis could significantly improve the clinical diagnosis of neurological disorders. %, including epilepsy.
 
%Although there is a wide variety of pathologies, conditions and network reorganisations that can lead to epilepsy, there is one observation that is crucial to its detection: the physiological signature of different seizures shows remarkable similarities from case to case.  Specifically, a number of complex morphological patterns can be identified at the onset of a seizure by monitoring the brain activity signal \cite{PDG13}. However, t
Detecting seizures using EEG or iEEG recordings presents several challenges. Firstly, distinguishing seizure activity from background noise or artifacts is often difficult, especially in long-term monitoring where false positives can occur. Secondly, the variability in seizure presentation between patients adds complexity, as there is no universal pattern or signature that reliably indicates seizure onset \cite{gotman1990automatic}. Additionally, the low signal-to-noise ratio of EEG or iEEG recordings can obscure subtle seizure activity, resulting in missed detections or false negatives. Moreover, interpreting EEG or iEEG data requires specialized training, making automated seizure detection algorithms prone to error without proper validation and refinement. Lastly, the computational complexity of analyzing large amounts of EEG or iEEG data in real-time presents a significant logistical challenge, requiring the use of efficient algorithms and computational resources for practical implementation \cite{karoly2017circadian}. Addressing these challenges is crucial for improving the accuracy and reliability of seizure detection methods in clinical practice. One promising approach to tackling these issues is to study the underlying nonlinear dynamics present in EEG/iEEG signals.

Epileptic seizures, as well as broader brain dynamics, can be understood through the lens of nonlinear dynamical systems \cite{SohanianHaghighi2017, Jirsa14}. The groundbreaking work of Hodgkin and Huxley, recipients of the Nobel Prize in 1963, demonstrated that neuron activity can be effectively modeled using a set of nonlinear differential equations \cite{hodgkin1952quantitative}. This discovery holds particular significance in the study of epileptic seizures, which are characterized by a synchronization in the brain network, replacing the usual complex nature of neuronal activity with behavior closer to chaos.
%The brain comprises intricate networks of billions of interconnected neurons, within which nonlinear interactions are ubiquitous. Individual neurons dynamically and nonlinearly influence one another, giving rise to emergent phenomena where intricate patterns of activity emerge from neural circuits \cite{SohanianHaghighi2017, Jirsa14}. Seizure activity exemplifies such emergent behavior, characterized by abnormal synchronization and oscillations spanning large brain regions. 
Concretely, seizures signify a shift from typical brain function to a pathological state marked by hypersynchronous and hyperexcitable neural activity. This transition often hinges on nonlinear phenomena, such as bifurcations or phase transitions, where small changes in neuronal dynamics trigger abrupt shifts in brain states \cite{sackellares2000epilepsy}.
%The study of these nonlinear features and their transitions offers profound insights into the mechanisms underlying seizure generation and propagation. 
Various approaches have been employed to detect seizures using nonlinear dynamics tools in epilepsy research. These range from entropy-based methods \cite{veisi2007fast} to Liapunov analysis \cite{khoa2012detecting} and Kolmogorov complexity \cite{radhakrishnan1998estimating}. However, many of these metrics focus on isolated signals, thereby overlooking some multidimensional aspects of the system. Additionally, these analyses are often susceptible to noise, requiring  signal pre-processing and making the implementation of real-time algorithms more challenging.
  
%In the search for improved seizure detection methods, researchers often struggle with the unavailability of the underlying equations describing neural dynamical systems. 
In practice, neural dynamical systems are not in general explicitly described, e.g. by a system of differential equations, but rather by a collection of concrete observations. Consequently, the analysis of neural dynamics often involves studying the topology of their \textit{attractors} --- invariant subsets of phase space towards which the system tends to evolve \cite{BW83, S67, GL02}. In the context of epilepsy, EEG/iEEG and MEG signals serve as measures of certain variables associated with the underlying neural dynamical system. The simultaneous embedding of these signals in a high-dimensional Euclidean space offers insights into the trajectories that approximate the associated attractor. Furthermore, the consistency of the dynamical properties observed in most spontaneous and evoked seizures across brain regions and species \cite{PDG13} suggests the existence of \textit{intrinsic invariant geometric properties} in the embedding of the neurophysiological signals, which encode robust information from the underlying dynamics.

Recently, the use of algebraic topology in the analysis of a wide variety of unstructured data has been remarkably successful. Topological Data Analysis (TDA) has emerged as a new mathematical field that aims to provide sound algebraic, statistical and algorithmic methods to infer, analyse and exploit the complex topological and geometric structures underlying raw data. Persistent homology \cite{ELZ02, Ghrist08, zomorodian_2005}, the vanguard technique in TDA, has shown successful results in the analysis and development of new theories in an increasing number of scientific fields such as medicine \cite{nicolau2011topology}, biology \cite{yao2009topological}, genomics \cite{rabadan2019topological}, physics \cite{cole2021quantitative}, chemistry \cite{lee2017quantifying}, among others. In the particular field of multivariate time series analysis, information about quasi-periodic patterns and changes in dynamics can be robustly detected in terms of the (persistent) homology of the embeddings \cite{PH15, P19, FBMG21}. Some initial work on the application of topological invariants to the detection of epileptic signals has been developed \cite{PRTM18, M16}. However, these methods require complex preprocessing of the whole data and are not applicable to real-time analysis

In this work, we propose a novel technique for real-time tracking and detection of epileptic seizures from raw neurophysiological recordings. The key to our method is the identification of topological biomarkers that successfully capture the changes in brain dynamics associated with epilepsy.
The time-evolving behaviour of brain activity is geometrically encoded in terms of the \textit{sliding-window embedding} of the multivariate time series derived from the simultaneously recorded channels.
We extract topological features from the signals by computing the persistent homology of the time-varying point clouds obtained from the sequential embeddings in a high-dimensional Euclidean space.
The evolution of the dynamics is represented in a path of \textit{persistence diagrams} --- the output of the persistence homology calculation --- whose approximate first derivative quantifies the different states of brain activity. Specifically, the values of the first derivative correspond to the amount of change in global neural activity, with high values indicating the start and end points of an epileptic seizure. The same procedure applied to a discriminative sliding window embedding for each channel (using the theory of Takens' delay embedding \cite{Tak81}) is able to capture the specific channel(s) where the seizure starts and/or generalises).
We also show that the \textit{total persistence} \cite{ST20} --- a numerical topological summary computed from persistence diagrams --- is sensitive to the presence of an epileptic seizure.

Our approach provides a scalable computational method for tracking the simultaneous evolution of raw neurophysiological recordings in real time, which is robust to noise due to the \textit{stability} property of persistent homology \cite{cohen2005stability}. We present the results of our methods on patients in a wide range of clinical situations --- from focal to global seizures --- and types of neurophysiological data --- from scalp and intracranial EEG to MEG. Remarkably, we achieve high accuracy in detecting ictal states under different conditions, demonstrating a robust mechanism for detecting the intrinsic signature present in seizure dynamics.
This finding has great potential for future applications in seizure control and neuromodulation. By integrating our method with \textit{closed-loop algorithms}, we could contribute to the design of electrical deep brain stimulation (DBS) systems capable of disrupting epileptic behaviour. These systems  would include real-time monitoring of symptom \textit{topological} biomarkers tailored to each patient, to detect seizures and automatically intervene when necessary.
%%%%%%%%%%%%%%%%%%%%%%%%%%

\section{Methods:\\ topological analysis of physiological recordings} 

We propose a method to analyze physiological recordings using tools from algebraic topology and dynamics. Although these techniques are primarily developed to detect changes in dynamics and synchronization from multichannel recordings of brain activity, they could also be applied to various other physiological signals and systems.

In the next sections we will see how to associate point clouds to collections of real-valued time series obtained from multichannel EEG/iEEG and MEG recordings, to subsequently apply a topological pipeline to extract persistence biomarkers
(Figure \ref{fig:persistent_homology}).
We start with a brief review of the main concepts and properties of persistent homology, the central theory in Topological Data Analysis, that we will employ later in our method.

%-------------------------------------------------------------------------------------------------------
\subsection{Persistent Homology} 
One of the best known computable invariants in algebraic topology is \textit{homology} \cite{Hatcher02}. This invariant assigns to every topological space $X$ a sequence $H_i(X)$ of groups that captures geometric signatures at dimension $i$. Indeed, the rank of the group  $H_i(X)$ defines the \textit{$i$-th Betti number} $\beta_i$ of $X$ and it quantifies the number of (independent) $i$-dimensional holes existing in $X$. For instance, $\beta_0$ represents the number of connected components, $\beta_1$ the number of 1-dimensional cycles, $\beta_2$ the number of voids and so on.
There exist efficient algorithms---based on normal matrix decompositions---to compute homology from the combinatorial descriptions of topological spaces.
Among the most used discrete representations are the \textit{simplicial complexes}, structures composed by simplices of different dimension (such as points, edges, triangles, etc) that encode all the relevant the topological information of a space (for instance, a sphere can be codified as the boundary of a tetrahedron).

\begin{figure}[htb!]
    \centering
    \includegraphics[width=\textwidth]{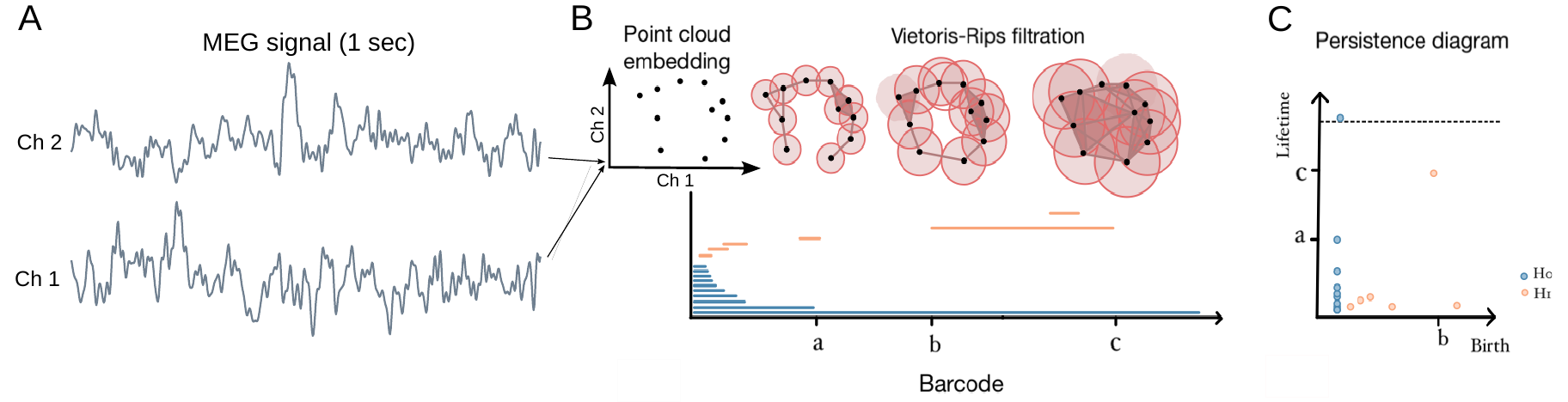}
    \caption{\textbf{Persistent Homology Pipeline.} \textbf{A.}  Example of two-channels MEG recording. \textbf{B.} Point cloud embedding of the signals and the Vietoris-Rips filtration (top). Barcode description of the persistent homology (botton). \textbf{C.} Persistence diagram. Every point represents a generator in homology, the first coordinate represents its birth and the second coordinate, its lifetime.} 
    \label{fig:persistent_homology}
\end{figure}

When analysing data, the topological space is often described through a finite sample of its points. In that case, the standard algorithm to compute homology is unable to recover any relevant property of the topology of the underlying structure, since it literally interprets the input as a finite discrete set.
Persistent homology \cite{ELZ02, Ghrist08, zomorodian_2005}  overcomes this limitation by looking at the sample at different scales of spatial resolution. More concretely, this computational method quantifies the evolution of homology groups of the point cloud when replacing each point by a ball of common increasing radius. Given a point cloud $X_n$ of size $n$ embedded in a Euclidean space, the \textit{Vietoris-Rips complex} $\VR_\epsilon$ at scale $\epsilon>0$ is a simplicial complex  with a simplex of dimension $m$ of for every subset of data points $\{x_0, x_1,\dots, x_m\}$ at pairwise distances less than $\epsilon$. Geometrically, $\VR_\epsilon$ is built as follows. Construct a ball of radius $\epsilon/2$ with center on each data point. Then, put an edge joining the center of every pair of intersecting balls. Finally, every time there is a subset of $m+1$ points pairwise connected with edges, then fill with the convex hull of the points in $\RR^{2m+1}$ (these are the $m$-simplices) (see Figure \ref{fig:persistent_homology}B). Notice that if $\epsilon<\epsilon'$, then $\VR_\epsilon\subseteq \VR_{\epsilon'}$. So $\{\VR_\epsilon\}_{\epsilon>0}$  represents a filtration of simplicial complexes whose homology groups capture the evolution of the topology of the point cloud for different resolution scales. The persistent homology of $X_n$ \textit{at degree $k$} --- which infers the holes of dimension $k$ ---  is the family of homology groups $\{H_k(\VR_\epsilon)\}_{\epsilon>0}$, along with the induced  maps $H_k(\VR_\epsilon)\to H_k(\VR_{\epsilon'})$ every time $\epsilon<\epsilon'$. All this information can be summarized as the list of parameter scales at which every generator in homology appears (birth) and disappears (death) (see the \textit{barcode} at Figure \ref{fig:persistent_homology}B). This suggests a graphical representation of the  persistent homology at degree $k$ as a diagram called \textit{persistence diagram} --- denoted by $\dgm_k(X_n)$ --- that represents every generator as a two-coordinated point whose first coordinate means its birth and the second coordinate, its death (see Figure \ref{fig:persistent_homology}C). For technical reasons, the diagonal (with generators with the same birth as death) will also be considered as part of every persistence diagram, with infinite multiplicity.

Similarity and differences in the topological signatures captured by persistence diagrams can be measured, e.g. with the Wasserstein distance. Given $\D$ and $\D'$ persistence diagrams and $p\geq 1$, its \textit{$p$-Wasserstein distance} $d_{W,p}$  is defined as
\[d_{W,p}(\D, \D') = \inf_{\psi:\D \to \D'} \left(\sum_{(x,y)\in \D}||(x,y)-\psi(x,y)||^p_{\infty}\right)^{1/p}\] 
where the infimum is taken over all bijections $\psi$ between generators in $\D$ and $\D'$  with finite death. The Wasserstein distance only captures the fundamental differences in the underlying topology of point clouds, whereas small changes in the input data are quantified as only a small change in the associated persistence diagram \cite{EH10}. This property of persistent homology is known as \textit{stability}, and makes it a useful tool for dealing with real-world, noisy data.

%%%%% -------------------------------- Figure 1 ------------------------------------------------

%%%---------------------------------------------------------------------------------------------------------
\subsection{Topology of time series} 
Let $\varphi_1, \varphi_2, \dots, \varphi_n:[0,T]\to \RR$ be a collection of real-valued time series, such as multichannel EEG and MEG recordings.
We will assume that the time series $\{\varphi_i\}_{1\leq i \leq n}$ are the result of observation functions applied to an underlying dynamical system. Namely, we assume that there exist a global continuous-time  dynamical system $(X, \phi)$ (where $X$ is the phase space and $\phi \colon \mathbb{R} \times X \to X$ is the evolution function), observation maps $\{F_i:X\to \mathbb R\}_{1\leq i \leq n}$  and an initial state $x\in X$ that produce the time series
\begin{align*}\varphi_i:\mathbb{R}&\to \mathbb{R}\\
							t&\mapsto F_i(\phi(t,x))
							\end{align*}
for all $1\leq i \leq n$.

We will analyze the underlying (unknown) dynamical system by studying the topology of their attractors \cite{A96, AB93,  S67, Tak81, W74}. 
In practice, we  estimate the attractor $M$ towards which the system is converging as the  embedding of the time series $\varphi = \{\varphi_i\}_{1\leq i \leq n}$ in $\RR^n$, defined as  \[\mathrm{E}_\varphi = 
\{(\varphi_1(t), \varphi_2(t), \dots, \varphi_n(t)): t\in [0,T]\}.\]

\subsection{Sliding-Window Embedding} Epileptic seizures are characterized by the abrupt switching between different states in the brain dynamics. Such transitions are reflected in changes in the topology of the local embeddings of the dynamic observations.
Given a collection of real-valued time series $\varphi= \{\varphi_i\}_{1\leq i \leq n}$ and a window-size $0<w\leq T$, we define the \textit{sliding-window embedding} (SWE) at $t\in [w, T]$ as the subset of $\RR^n$ given by the topological space\[\SWE_{\varphi, w}(t) = \{(\varphi_1(s), \varphi_2(s), \dots, \varphi_n(s)): s \in [t-w, t]\}.\]
This procedure describes a \textit{path $\gamma(t)$ of time-evolving topological spaces}:
\begin{align*}
\gamma: [w,T]&\to \mathrm{\bf Top}\\
t&\mapsto \mathrm{SWE}_{\varphi,w}(t).
\end{align*}

Notice that, in practical applications, the signals are only recorded at a finite set of equidistant sample times $t_1, t_2, \dots, t_m \in [0,T]$. Hence, the topological spaces $\mathrm{SWE}_{\varphi,w}(t)$ 
are described by just a finite sample of its points. Computational methods for the inference of topological invariants from point clouds, such as \textit{persistence homology}, will be then employed.

\subsection{Persistence summaries}\label{topological summaries} Let $\varphi= \{\varphi_i\}_{1\leq i \leq n}$ be a collection of real-valued time series on $[0,T]$ and let $w$ be the window size.  We will capture information of the evolution of the topology of the spaces $\mathrm{SWE}_{\varphi, w}(t)$ using persistent homology. Concretely, we consider the path of time-varying persistence diagrams
\begin{align}
PH_k:[w,T]&\to \mathrm{\bf Dgm} \nonumber\\
t&\mapsto \mathrm{dgm}_k(\SWE_{\varphi,w}(t)). \label{path_dgm}
\end{align}

In practice, we have
$\{t_j\}_{1\leq j \leq m}$ be a finite sample of equidistant times and set $0<w\leq T$ as a multiple of $t_j-t_{j-1}$.
In order to quantify the local changes in the topology of the embedding along the time interval, we estimate the \textit{first order derivative} $\frac{\partial PH_k}{\partial{t}}$ of the map $PH_k(t)$ \eqref{path_dgm} as the difference quotient
\begin{equation}
\dfrac{d_{W,p}\Big(\dgm_k\big(\SWE_{\varphi, w}(t_j)), \dgm_k(\SWE_{\varphi, w}(t_{j-1})\big)\Big)}{t_{j}-t_{j-1}}
\label{derivative}
\end{equation}
for every $1\leq j \leq m-1$ (c.f. \cite[Section 4]{FBMG21}).

We will also associate to every diagram  $PH_k(t_j)$, $1\leq j \leq m$, a measure of its \textit{topological complexity}.  Given $\D$ a persistence diagram, the \textit{persistence} of a  generator $(x,y)\in \D$ with finite death is  $\pers(x,y) := y-x$.
For $k\geq 1$, the \textit{total persistence} of $\D$ \cite{CEHM10} is defined as 
\[\Pers(\mathcal{D}) = %\left(
\sum_{(x,y) \in \mathcal{D}} \pers(x,y)
%\right)
.\]
%%%-------------------------- Figure 2 ----------------------------------------------------
\begin{figure}[htb!]
    \centering
    \includegraphics[width=\textwidth]{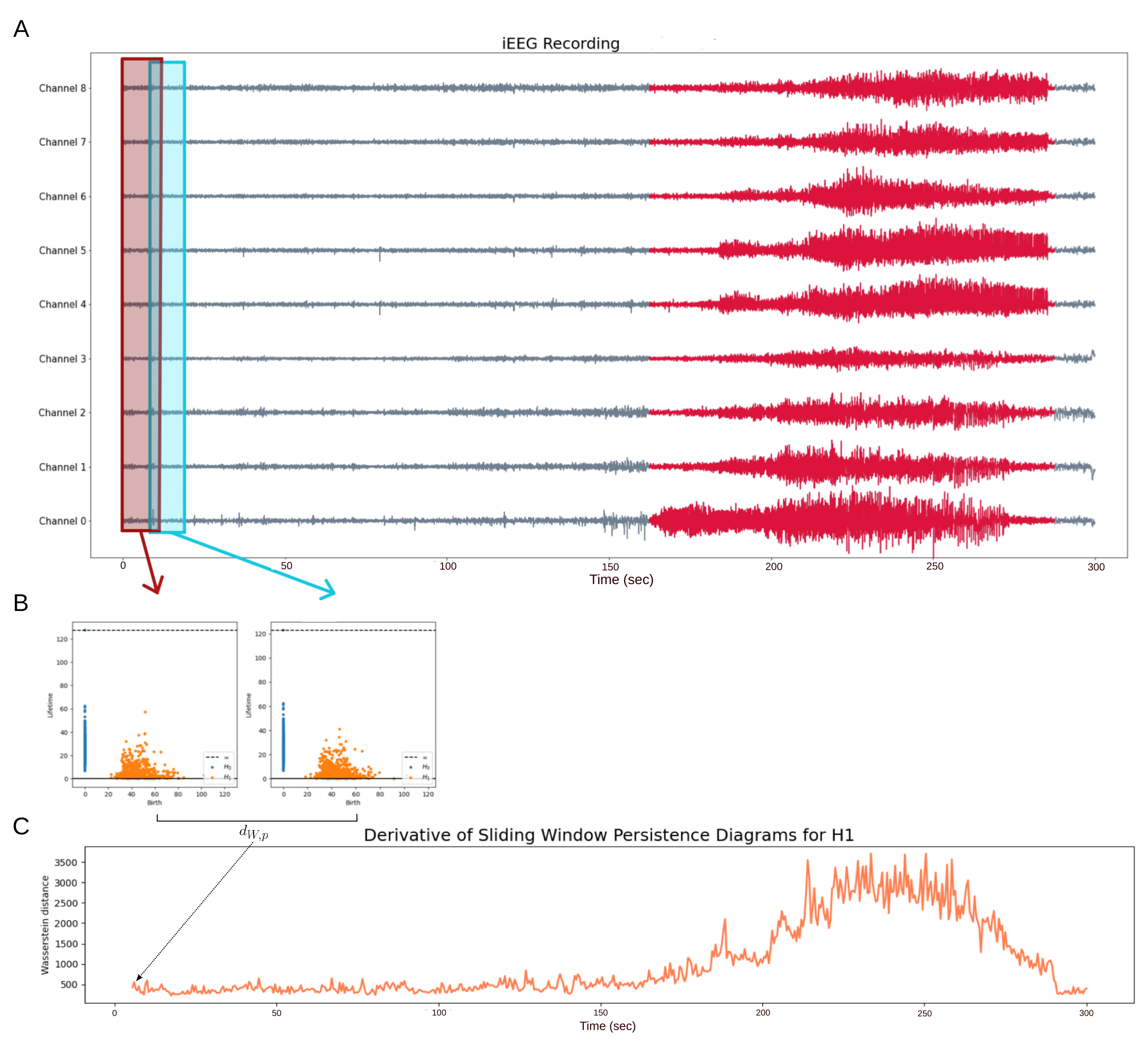}
    \caption{\textbf{Topological biomarker (derivative).} Pipeline for the computation of the estimator of the derivative of the path of persistence diagrams. \textbf{A.} Sliding window embedding. \textbf{B.} Path of persistence diagrams. \textbf{C.} Estimator of the first order derivative using Wasserstein distance of diagrams.}
    \label{fig:path_persistent_diagrams}
\end{figure}

%--------------------------------------------------------------------

\subsection{Global vs local dynamics and delay-embeddings} \label{takens}
%\textcolor{red}{Epileptic seizures may involve either part of the cerebral hemisphere (\textit{focal} seizures) or, on the contrary, the entire brain (\textit{generalized} seizures)---the latter usually presenting loss of consciousness. This difference is reflected in either a local change in the brain dynamics, in which only a fraction of the set of the physiological signals presents a different behaviour, or a global change, in which all the signals present a synchronized seizure pattern.}

Global versus local dynamics can be geometrically understood using algebraic and differential topology. 
Indeed, time-delay embeddings of scalar time-series data, introduced by Takens in the 80's \cite{Tak81}, is a well-known technique to recover the underlying global dynamics induced by a single observation of the system. Concretely, Takens' theorem gives conditions under which the topology of a smooth attractor can be reconstructed from a generic observable function. It implies in particular that if $\varphi_i(t)$ is a real valued signal (for instance, one channel of a multichannel recording), then the \textit{delay coordinate map}
\[
t\mapsto \Big(\varphi_i(t), \varphi_i(t+\tau), \varphi_i(t+2\tau) \dots, \varphi_i(t+(D-1)\tau)\Big)
\]
is an embedding of an orbit. Here $D$ is the embedding dimension (theoretically representing the number of variables of the original system) and $\tau$ is the time delay.
 If the attractor is a smooth manifold $M$ of dimension $d$, under certain conditions Takens' theorem  implies that the delay embedding of the signal with $D\geq 2d+1$ is  diffeomorphic to $M$ (here, the dimensional bounds are related to those of the Whitney Embedding Theorem \cite{whitney1936differentiable}).

Time-delay embeddings allow us to develop a technique for understanding global versus local dynamics. Specifically, we analyze the dynamics induced by each individual signal from a multichannel electrophysiological recording to determine their consistency. If different dynamics coexist (that is, the signals represent different dynamics), then different attractors should arise.
This is the case of focal seizures, where distinct dynamics occur in different brain areas—some reflecting normal behavior and others reflecting epileptic behavior. In contrast, for generalized seizures, the global dynamics induced by every channel should align.

The procedure works as follows. For every signal $\varphi_i$ ($1\leq i \leq n$) from the multichannel recording, compute the \textit{sliding-window time-delay embedding} at $t\in [w, T-(D-1)\tau]$, defined as the subset of $\RR^D$ given by
\begin{equation}\mathrm{SWE}_{\varphi_i, w}(t) = \{(\varphi_i(s), \varphi_i(s+\tau), \dots, \varphi_{D-1}(s+(D-1)\tau)): s \in [t-w, t]\}.
\label{takens window delay embedding}
\end{equation} Then, the analysis continues as in section \hyperref[topological summaries]{Topological summaries}. Namely,  compute for every signal $\varphi_i$, the approximate of the \textit{first order derivative} of the map \eqref{path_dgm}.

\section{Data}\label{data} In this study, we  analyze three different databases.

The first database was prepared by the Toronto Western Hospital, Canada, and it contains both intracranial EEG (iEEG) and MEG recordings from four patients with a wide range of clinical situations. This data from Patient 1 was previously analyzed  in \cite{V11}, while Patients 2-4 were investigated in \cite{D05}.

\begin{itemize}
\item  \textbf{Patient 1.} %(CC).
It presents a medically refractory temporal lobe epilepsy. The iEEG recording was obtained as part of the patient’s routine clinical presurgical analysis. The electrodes were positioned in a number of locations including the amygdala and hippocampal structures of both temporal lobes. The data consists of 300 seconds of recording from 9 channels at 200 Hz sampling rate. %This data also was previously analyzed  in \cite{V11}. 
%\textcolor{red}{Seizure starts at $t_0=30000$ and ends at $t_1=58000$}
\item \textbf{Patient 2.} %(RAO).
It presents a symptomatic (formerly known as secondary) generalized epilepsy. The data consists of a MEG recording of 120 seconds from 171 channels at 625 Hz sampling rate. The recording sensors covered the whole cerebral cortex. %This data was previously analyzed in \cite{D05}.%\textcolor{red}{The seizure starts at $t_0=28375$ and ends at $t_1=37500$.}

\item \textbf{Patient 3. }%(MAC)  
It presents a primary generalized absence epilepsy. The data consists of a MEG recording of 120 seconds from 144 channels at 625 Hz sampling rate. The recording sensors also covered the whole cerebral cortex. %This data was previously analyzed in \cite{D05}. 
%\textcolor{red}{Seizure starts at $t_0=60000$ and ends at $t_1=68125$}

\item \textbf{Patient 4.} %(Atkar)  
It has a frontal lobe epilepsy with a right frontal subcortical \textit{heterotopia} (a form of cortical dysplasia). The data consists of a MEG recording of 120 seconds from 148 channels at 625 Hz sampling rate. The recording sensors also covered the whole cerebral cortex.%\textcolor{red}{Seizure starts at $t_0 = 59000$ and ends at $t_1=67600$.}

\end{itemize}
The second database is the open-source scalp EEG database CHB-MIT \cite{CHB-MIT}, provided by the Massachusetts Institute of Technology, USA. The recordings were collected from pediatric subjects with intractable seizures at the Children’s Hospital Boston. We selected for this exposition a particular patient that 
presents complex seizure information, which turned out to be hard to identify using standard methods.

\begin{itemize}
\item \textbf{Patient 5.} %(CHB-MIT)}.
It corresponds to case \textit{chb01} in the CHB-MIT database. The data consists of a EEG with  23 sensors in a 10/20 system with 256 Hz sampling frequency.
%During these recordings, the patient experienced seven seizures.
\end{itemize}

%\textcolor{teal}{The third database utilized in this study is the open-source EEG database \cite{nasreddine2021epileptic} comprising recordings from patients at the Epilepsy Monitoring Unit of the American University of Beirut Medical Center
%Specifically, the data were derived from patients diagnosed with focal epilepsy who underwent pre-surgical evaluation with long-term video-EEG monitoring to assess their eligibility for epilepsy surgery. During this evaluation process, antiepileptic drugs were discontinued to facilitate recording of the patients' habitual seizures.
%In total, the dataset includes information from six patients, covering a total of 20 recording sessions. The recorded data represent measurements from 21 scalp electrodes, adhering to the 10-20 electrode system, and were sampled at a frequency of 500 Hz. The available EEG data sets were pre-processed by band-pass filtering in the frequency range from 1/1.6 Hz to 70 Hz, excluding the 50 Hz power supply frequency.
%The EEG signals were segmented into 1-second acquisition windows and subsequently labelled as either normal or lesional data. Notably, the dataset includes several seizure types, including complex partial seizures (CPS), electrographic seizures (ES), and video-detected seizures with no discernible visual changes on the EEG. The distribution of acquisition windows across these conditions was as follows Normal (N) - 3895 epoch, complex partial seizures (CPS) - 3034 epoch, electrographic seizures (ES) - 705 epoch, video-detected seizures (VS) - 111 epoch.}

The third database utilized in this study is the open-source EEG database \cite{nasreddine2021epileptic}  comprising recordings from patients at the Epilepsy Monitoring Unit of the American University of Beirut Medical Center. Specifically, the data were derived from patients diagnosed with focal epilepsy who underwent pre-surgical evaluation with long-term video-EEG monitoring to assess their eligibility for epilepsy surgery. During the evaluation process, antiepileptic drugs were discontinued to facilitate recording of the patients' habitual seizures. The recorded data represent measurements from 21 scalp electrodes, adhering to the 10-20 electrode system, and were sampled at  frequency  500 Hz. The available EEG data sets were pre-processed by band-pass filtering in the frequency range from 1/1.6 Hz to 70 Hz, excluding the 50 Hz power supply frequency. Certain channels were excluded from specific recordings due to artifact constraints. For the purpose of this exposition, a specific patient was selected to exemplify a type of seizures that are not visually detectable from the EEG recording, but only from video recordings of physical symptoms.

\begin{itemize}
\item \textbf{Patient 6.}
It corresponds to \textit{Individual 13, Recordings 0-2} (9 years, masculine) of the database \cite{nasreddine2021epileptic}. The data consists of  EEG recordings with 19 sensors (channels `EEG Cz-Ref' and `EEG Pz-Ref' omitted due to artifact constrains) at 500 Hz sampling frequency. 
%During these recordings, the patient experienced four seizures. 
The seizures from this patient are categorized as \textit{video detected}, with no visual changes over the EEG. 
\end{itemize}

%======================================================================================================
\section{Results}

\subsection{Topological seizure detection} \label{top seizure detection}
%We analyze different types of neurophysiological recordings for patients with a variety of epileptic diagnosis. 
In this section, we demonstrate how persistence summaries computed from collections of real-valued time series can reliably detect the ictal state. Specifically, we showcase the topological analysis of various neurophysiological recordings obtained from six patients diagnosed with different types of epilepsy (refer to \hyperref[data]{Data} for further details).

The window size $w$ for the sliding-window embedding is set to a value equivalent to 1 second of recording for patients 2, 3, and 4 (with a sampling rate of 625 Hz) and for patient 6 (with a sampling rate of 500 Hz), and to 2 seconds for patients 1 and 5 (with sampling rates of 200 Hz and 256 Hz, respectively). For every case study, all the topological summaries are computed for degree 0 and 1. The computation of the approximation of the first derivative is performed using the Wasserstein distance for $p=1$. %The $k$-total persistence is evaluated for $k=1$. 
The finite sample of the time interval %$[0,T]$ 
for every recording is determined as evenly spaced times with a stride equivalent to $0.5$ seconds.

\begin{figure}
    \centering
    \includegraphics[width=0.9\textwidth]{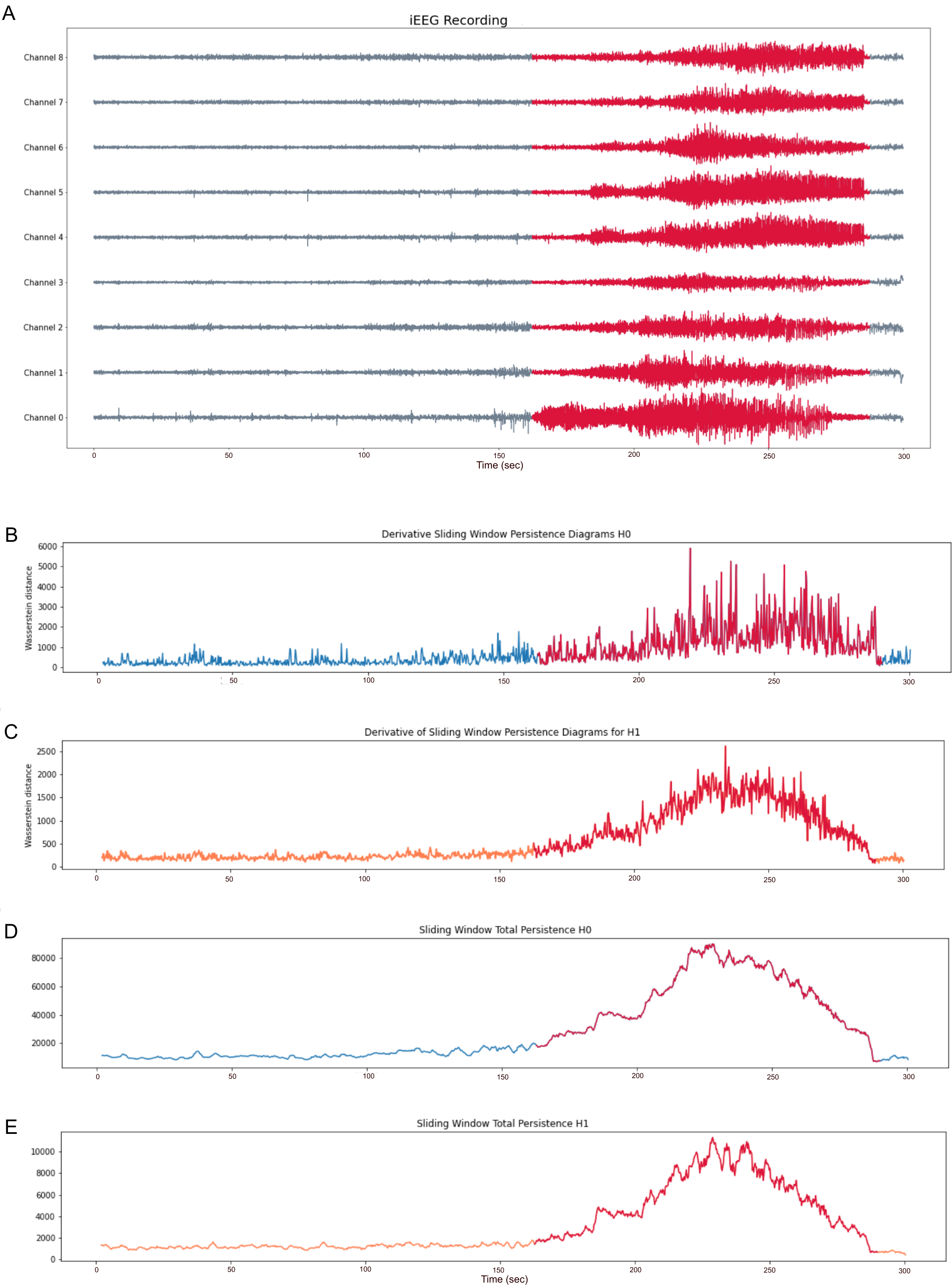}
    \caption{\textbf{Patient 1.} %CC
    The total time-interval of the analysis is $[0,300]$ sec, where the ictal state is at $[162, 290]$ sec (in red). \textbf{A.)} iEEG with 9 sensors at frequency 200 Hz. \textbf{B-C.)} Estimator of first derivative of the time-evolving persistence diagrams for degrees 0 and 1. \textbf{D-E.)} Total persistence for degrees 0 and 1.}
    \label{fig:patient_CC}
\end{figure}

%\subsubsection*{Results} 
Patient 1 presents a focal seizure that rapidly generalizes (see Figure \ref{fig:patient_CC}).
This is reflected in a gradual increase in the value of all the topological indicators, specially the ones associated to persistent homology at degree 1. On the other hand, the continuous decrease in the biomarkers by the end of the ictal state reflects the progressive global desynchronization.

\begin{figure}
    \centering
    \includegraphics[width=0.85\textwidth]{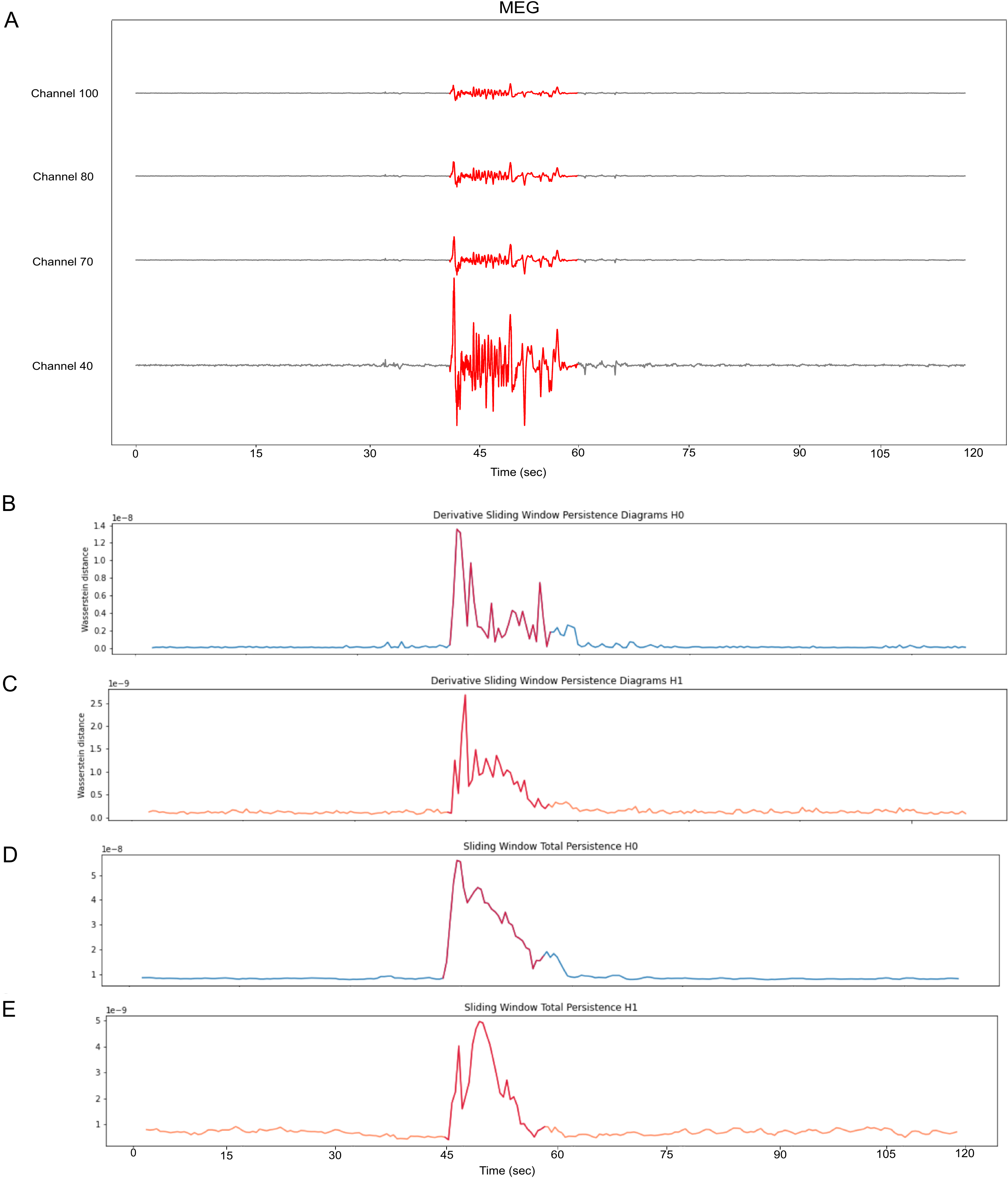}
    \caption{\textbf{Patient 2.} %RAO
    The total time-interval of the analysis is $[0,120]$ sec, where the ictal state is at $[45, 64]$ sec (in red).  \textbf{A.)} Raw signal of four of the 171 MEG signal at frequency 625 Hz. \textbf{B-C.)} Estimator of first derivative of the time-evolving persistence diagrams for degrees 0 and 1. \textbf{D-E.)} Total persistence for degrees 0 and 1.}
    \label{fig:patient_RAO}
\end{figure}

\begin{figure}
    \centering
    \includegraphics[width=0.8\textwidth]{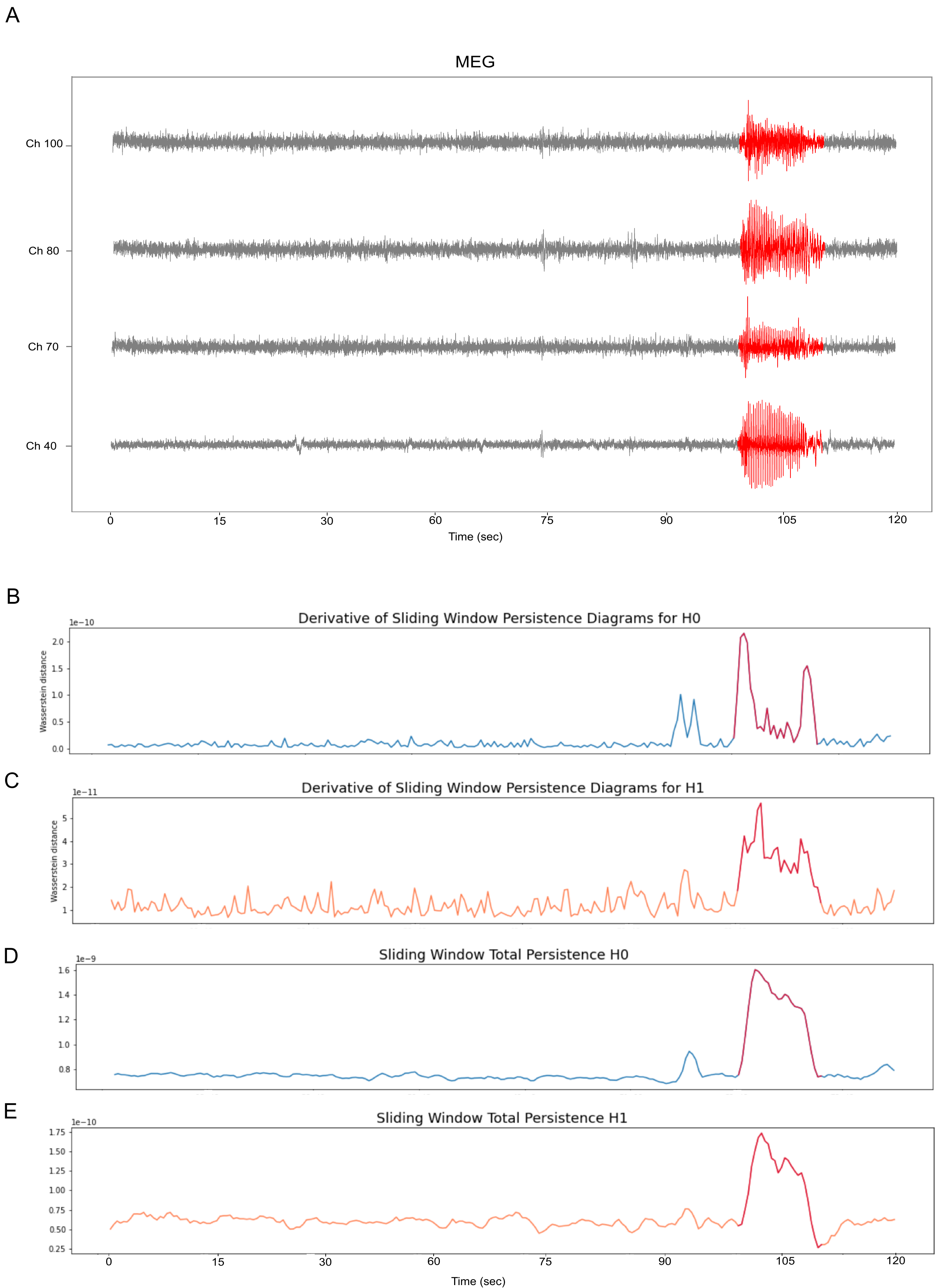}
    \caption{\textbf{Patient 3.} % MAC 
    The total time-interval of the analysis is $[0,120]$ sec, where the ictal state is at $[96, 109]$ sec (in red).  \textbf{A.} MEG  Raw signal of four of the  141 sensors  at frequency 625 Hz. \textbf{B-C.} Estimator of first derivative of the time-evolving persistence diagrams for degrees 0 and 1. \textbf{D-E.} Total persistence for degrees 0 and 1.}
    \label{fig:patient_MAC}
\end{figure}

Patients 2 and 3  present a global seizure (see Figures \ref{fig:patient_RAO} and \ref{fig:patient_MAC}). The onzet is determined by a synchronized spike that triggers all the topological indicators, with an abrupt decrease after the seizure ends. 
The peaks in the first derivative at the start and end of the ictal state reveals abrupt changes in the dynamics, whereas low values in the interior of the ictal state indicate a stable global dynamic during the seizure (see Subsection  \hyperref[interpretability]{Interpretability of the topological biomarkers}  for details on the interpretability of every topological indicator).

\begin{figure}
    \centering
    \includegraphics[width=0.85\textwidth]{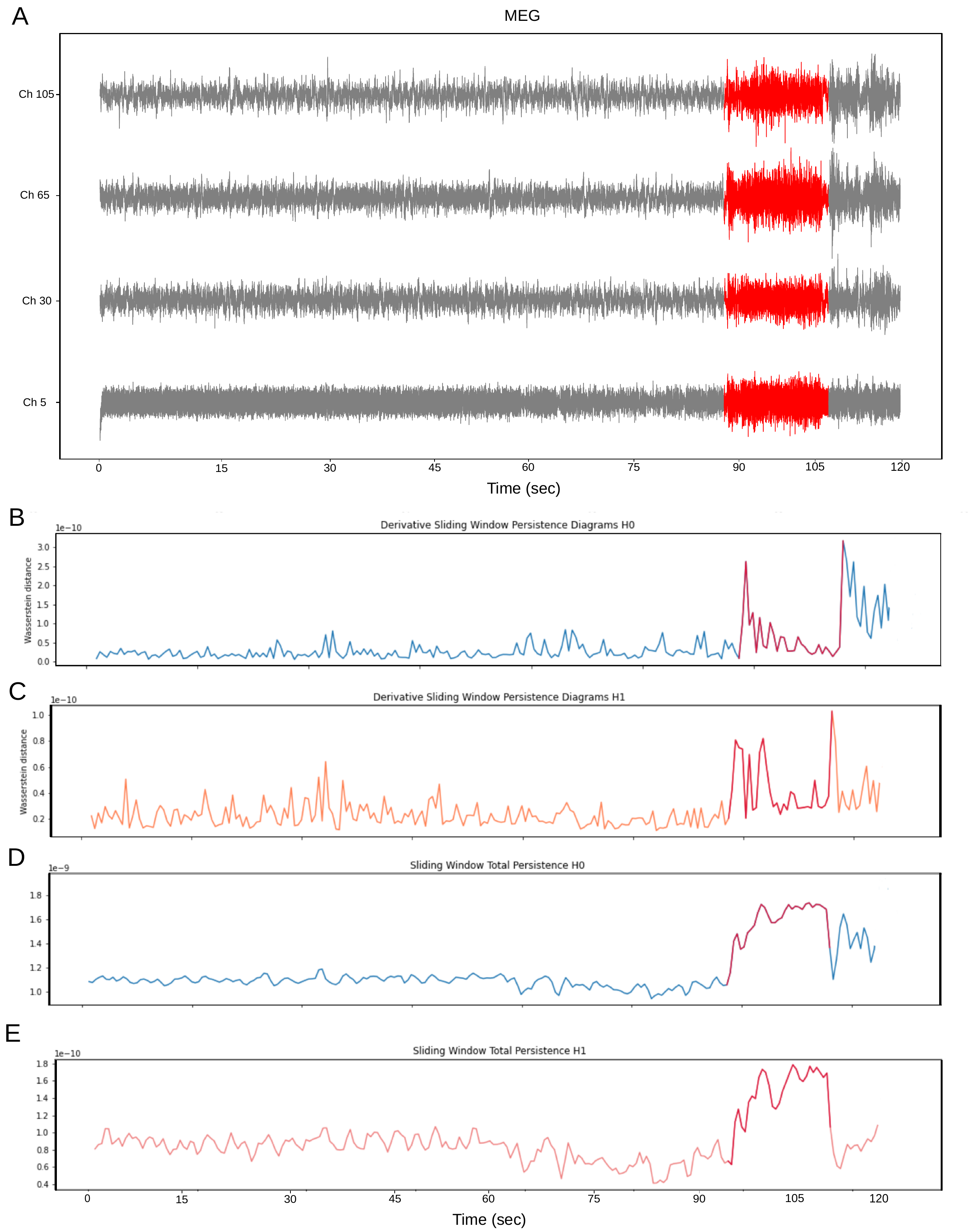}
    \caption{\textbf{Patient 4.} % Aktar.
    The total time-interval of the analysis is $[0,120]$ sec, where the ictal state is at $[94, 109]$ sec (in red). \textbf{A.} MEG with 148 sensors  at frequency 625 Hz. \textbf{B-C.} Estimator of first derivative of the time-evolving persistence diagrams for degrees 0 and 1. \textbf{D-E.} Total persistence for degrees 0 and 1.}
    \label{fig:patient_Aktar}
\end{figure}

Patient 4  presents a (focal) epileptic seizure in the fronto-temporal area, that is evidenced in a change of dynamics in only a subset of the channels in the MEG recording (see Figure \ref{fig:patient_Aktar}). 
As in the previous cases, there is a peak in the sliding-window derivatives for degree 0 and 1 at the beginning and end of the ictal state. However, this 
case presents a gradual decrease in the values of the derivative at the post ictal state, showing a slow return to baseline.
A similar situation can be read from the total persistence indicators, which present an abrupt increase during the ictal state in a plateau shape, and gradual recovery to baseline state after the seizure.

\begin{figure}
    \centering
    \includegraphics[width=0.9\textwidth]{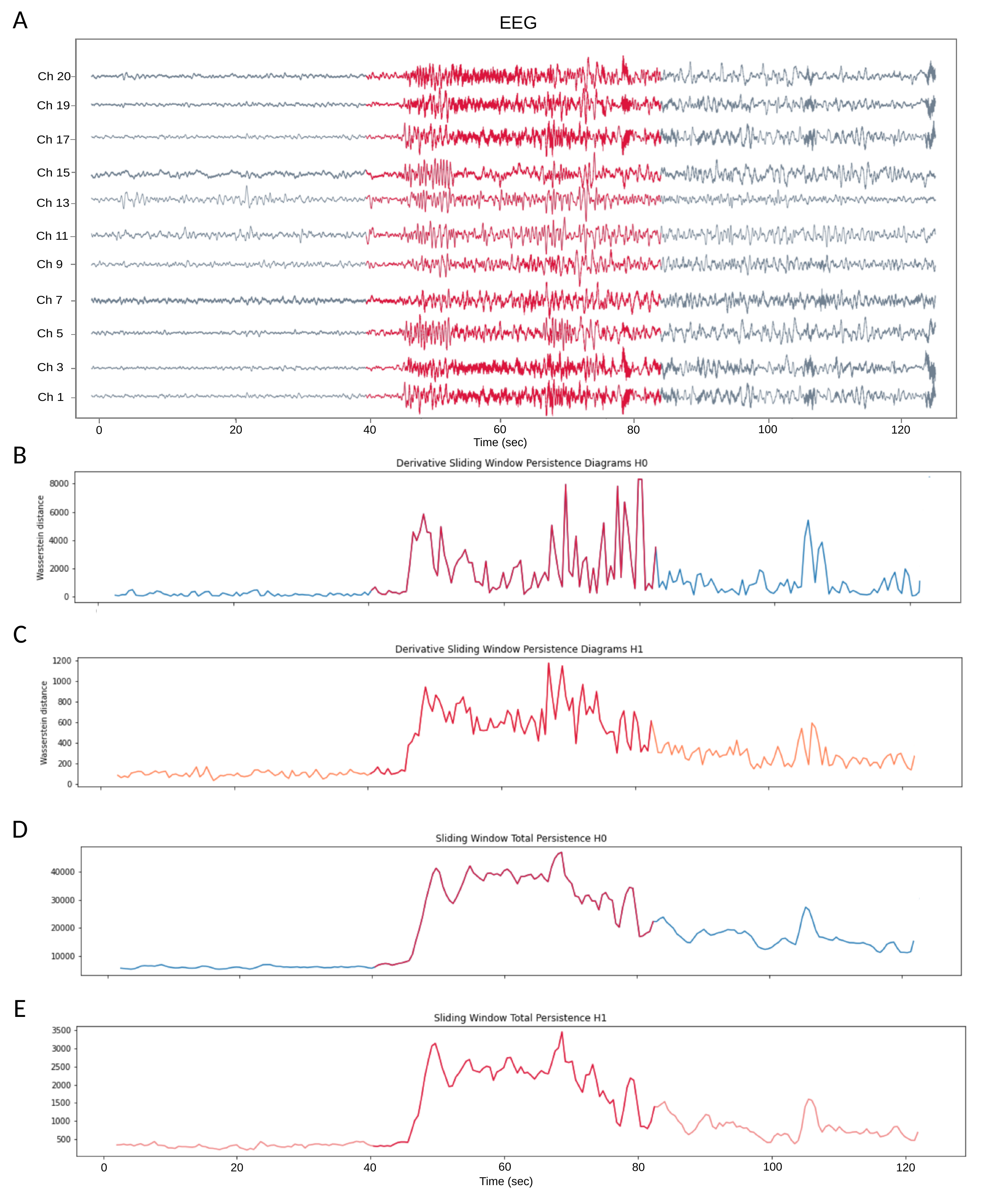}
    \caption{\textbf{Patient 5 (\textit{chb01} from CHB-MIT database \cite{CHB-MIT}).} The total time-interval of the analysis is $[0,120]$ sec, where the ictal state is at $[39, 81]$ sec (in red).  \textbf{A.} EEG with 23 sensors  at frequency 256 Hz. \textbf{B-C.} Estimator of first derivative of the time evolving persistence diagrams for degrees 0 and 1. \textbf{D-E.} Total persistence for degrees 0 and 1.}
    \label{fig:patient_CHB-MIT}
\end{figure}

Patient 5 presents a complex pattern during (and also after) the ictal state, that most standard techniques for seizure detection are not able to recognize (see \hyperref[comparison]{Comparison with standard seizure indicators}).
Remarkably, all our topological indicators are triggered during the ictal interval, whereas they present stable low values at the preictal stage (see Figure \ref{fig:patient_CHB-MIT}). They also show a gradual decrease during the postictal stage.

\begin{figure}
    \centering
    \includegraphics[width=0.75\textwidth, ]{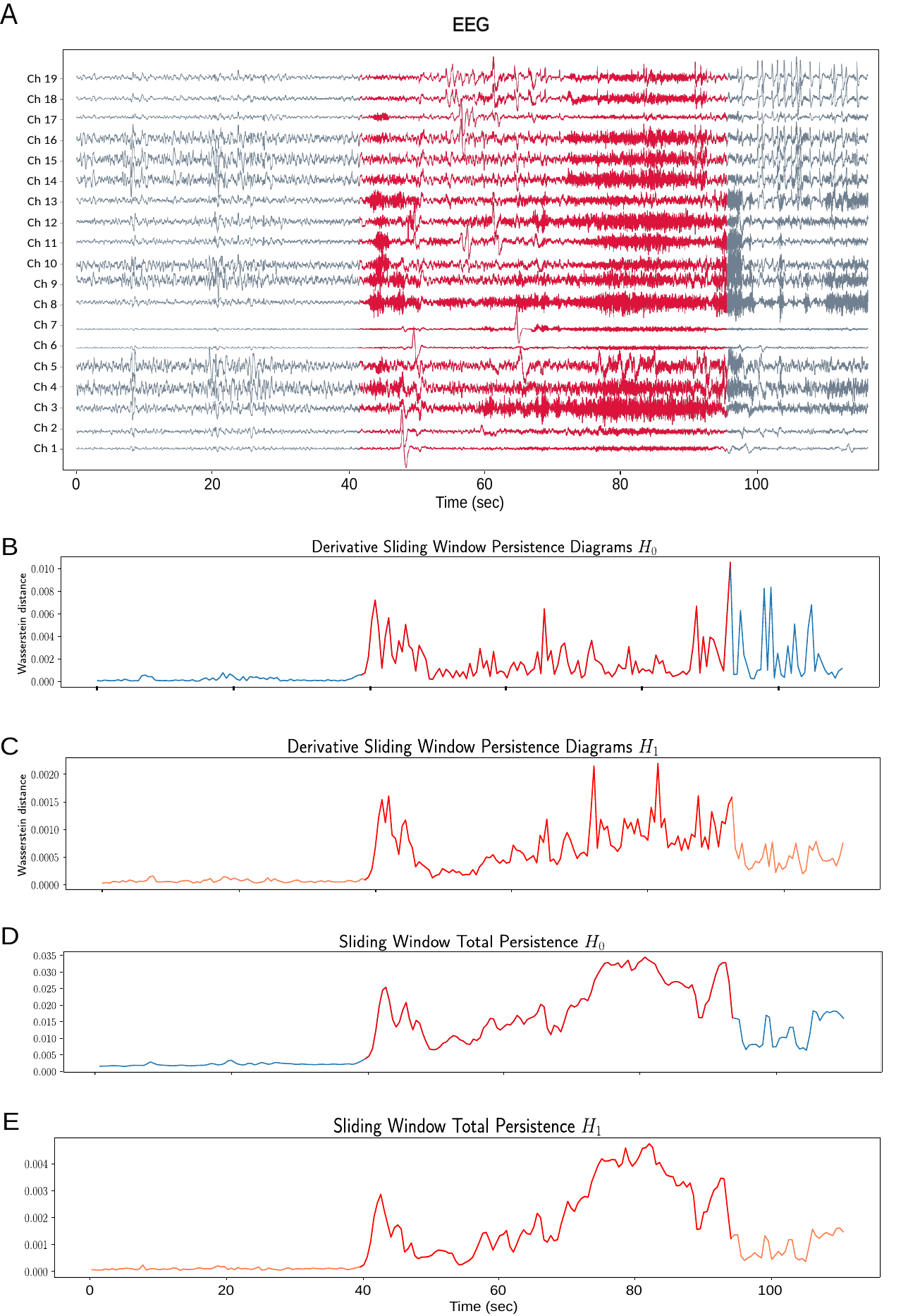}
    \caption{\textbf{Patient 6 (\textit{Individual 13, Record 1, Seizure 1} from Beirut database \cite{nasreddine2021epileptic}).} The total time-interval of the analysis is $110$ seconds, where the ictal state is at $[40,95]$ seconds (in red).  \textbf{A.} EEG with 19 sensors  at frequency 500 Hz. \textbf{B-C.} Estimator of first derivative of the time evolving persistence diagrams for degrees 0 and 1. \textbf{D-E.} Total persistence for degrees 0 and 1.}
    \label{fig:patient_Beirut}
\end{figure}

Patient 6 presents a type of seizure that was classified as \textit{video-detected with no discernible visual changes on the EEG}. However, our topological biomarkers from the EEG data clearly trigger during the seizure, while exhibiting flat behavior during the pre-ictal state. Although these biomarkers remain elevated after the marked termination point of the seizure onset, they gradually decrease in value (see Figure \ref{fig:patient_Beirut}), indicating complex dynamics during the post-ictal state. %\textcolor{red}{This case study demonstrates that the complexity of the dynamics captured by these topological biomarkers is more nuanced than just a binary classification, as they also quantify aspects of global seizure dynamics that are not discernible through standard signal inspection.}

%\begin{figure}
%    \centering   
%     \includesvg[width=0.89\textwidth]{figures/Beirut_13_EEG.svg}
%    \includegraphics[width=0.8\textwidth]{figures/SW_Beirut13_consecutive_distance_H0_250.png}
%    \includegraphics[width=0.8\textwidth]{figures/SW_Beirut13_consecutive_distance_H1_250.png}
%    \includegraphics[width=0.8\textwidth]{figures/SW_Beirut13_total_persistence_H0_250.png}
%    \includegraphics[width=0.8\textwidth]{figures/SW_Beirut13_total_persistence_H1_250.png}
%    \caption{\textbf{Patient 6 (\textit{Individual 13, Record 1, Seizure 1} from Beirut database \cite{nasreddine2021epileptic}).} The total time-interval of the analysis is $110$ seconds, where the ictal state is at $[40,95]$ seconds (in red).  \textbf{A.} EEG with 19 sensors  at frequency 500 Hz. \textbf{B-C.} Estimator of first derivative of the time evolving persistence diagrams for degrees 0 and 1. \textbf{D-E.} Total persistence for degrees 0 and 1.}
%    \label{fig:patient_Beirut}
%\end{figure}

\subsection{Interpretability of the topological biomarkers}\label{interpretability} In this section, we offer a detailed description of the connection between the seizure dynamics encoded by the physiological signals and the topological summaries.

\textit{Persistent homology at degree 0} quantifies the connectivity of the point cloud.
Every point at the persistent diagram $\D_0$ represents a generator (a connected component) that is born at the scale parameter $\epsilon = 0$ and dies at the value of $\epsilon>0$ at which the connected component represented by the generator joins another one in the Vietoris-Rips filtration.
In particular, there is always a single generator with infinite lifetime, indicating that the Vietoris-Rips complex for large values of $\epsilon$ is connected.
However, the persistence of the rest of the generators encodes relevant properties of the embedding of the signals. For instance, the persistence of the generator with maximum finite lifetime equals the diameter of the point cloud and, hence, it is correlated with the amplitude of the signals. Moreover, the variance of the values of the persistence of the generators in $\D_0$ is associated with the density of sampling of the underlying geometric space. Given that the size of the point cloud remains constant along all the time-varying embeddings, the distribution of the persistence of generators in $\D_0$ is related to the rate between amplitude (also spikes) and frequency, and the synchronization of the oscillations in all the signals simultaneously.

\textit{Persistent homology at degree 1} summarizes data about the (independent) 1-dimensional cycles that can be reconstructed from the point cloud. Generators with long lifetime represent structural cycles in the embedding, whereas the ones with short lifetime are associated with small holes in areas of low density. At the signal level, homology at degree 1 of the embedding encodes the presence of synchronized (quasi) periodic patterns, while its persistence amounts to its level of synchronization and its amplitude.

The \textit{estimator of the first derivative} of the persistence of the time-evolving point clouds measures the local changes in the geometry and topology of the embeddings. There is a number of situations quantified by this topological biomarker:
\begin{itemize}[label={--}]
\item abrupt change from low to high values of the derivative when the seizure starts and ends reflects a global change in the dynamics induced by a generalized seizure.
\item progressive increasing values of the derivative from low to high when the seizure starts and/or decreasing values when the seizure ends is an indicator of the starting of a focal seizure that generalizes by the end and, analogously, of brain activity that progressively desynchronizes by the end of the seizure.
\end{itemize}
Moreover, during the seizure the derivative might conserve high values (as at the start and end) or it might decrease the value of the highest peak (still having greater values than at the interictal state). This is consequence of the type of seizure pattern displayed at the ictal state:
\begin{itemize}[label={--}]
\item low values of the derivative after the seizure starts and before it ends means a consistent and uniform dynamic behaviour during the ictal state;%, until it ends when there is another peak showing a change in the dynamics;
\item high values of the derivative throughout the ictal state represents an non-stable seizure dynamic, in which there are constant changes in the amplitude and frequency, periodicity, spikes, and wave activity \cite{PDG13}.
\end{itemize}

The \textit{total persistence} biomarker consists of a summary of the lifetime of the generators in the time-evolving persistence diagrams, for every degree. It quantifies numerically the complexity of the dynamics encoded in the time-evolving embeddings by means of its topological features. During the seizure, the total persistence for both degrees 0 and 1 is triggered, while it presents low values during the interictal state (c.f. \hyperref[quantitative]{Statistical analysis of the total persistence biomarker}).

%Despite its complexity, seizure activity presents two major types of consistent patterns in neurophysiological recordings: fast oscillations and spikes with or without waves.

\subsection{Comparison with standard seizure indicators} \label{comparison}

In the study of electrophysiological signals, there is a rich variety of mathematical tools available for analysis. In particular, many of these tools have been used in the study of seizure detection \cite{harpale2016time, lehnertz2003seizure, nicolaou2012detection}. For this paper, we will compare our results with two analyses commonly used in the literature for signal processing. The first one is based on the time-frequency decomposition,  and the second is a non-linear measure derived from information theory.

\begin{figure}[htb!]
    \centering
    \includegraphics[width=1\textwidth]{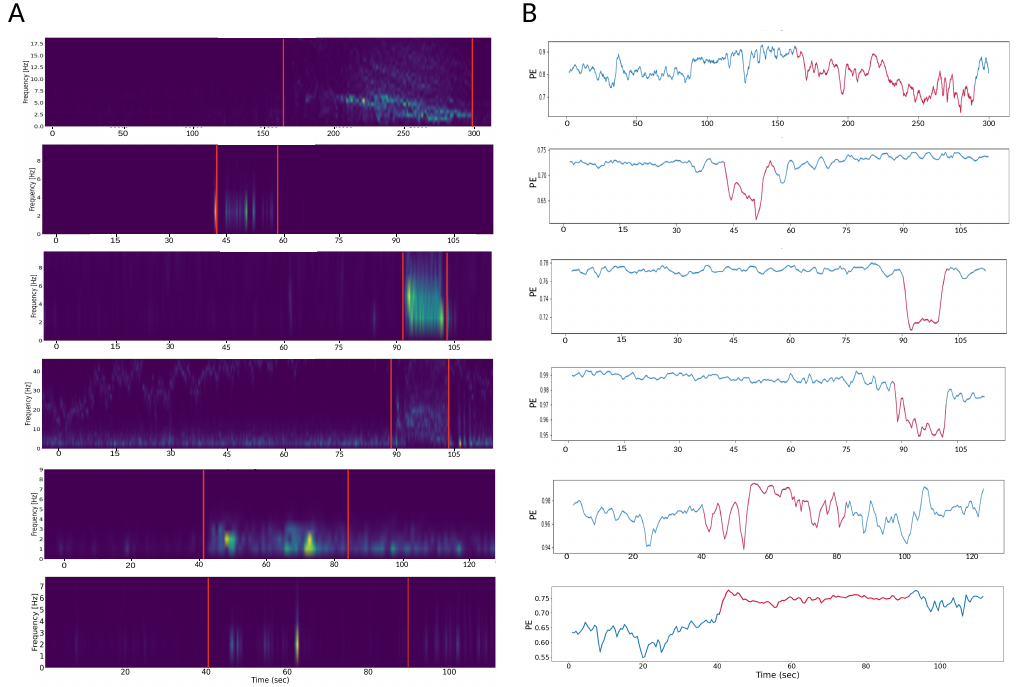}
    \caption{\textbf{Spectral and non-linear signal analysis.} Spectrograms associated to the Short-Time Fourier Transform (STFT) \textbf{(A)},  and averaged Permutation Entropy (PE) \textbf{(B)} from physiological recordings of Patients 1-6 (top to bottom). The ictal stages are indicated between the red bars and the red lines.}
    \label{fig:comp}
\end{figure}

%STFT
One of the most standard techniques in signal processing is the \textit{time-frequency analysis}, which decomposes a signal in both time and frequency domains simultaneously 
%The first tool is based on a time-frequency domain decomposition, which allows signals to be mapped into a two-dimensional function of frequency and time 
\cite{mustafa2011analysis}. The Short-Time Fourier Transform (STFT) is used to segment a signal into short-term segments by shifting the time window with some overlap \cite{zabidi2012short} (a process known as \textit{windowing}).  In our analysis, we computed the STFT for each channel of the recordings to obtain the time-frequency matrix for each channel (also known as \textit{spectrogram}). Finally, we calculated the mean of all the matrices across the channels, resulting in a single matrix representing the entire system (see Figure \ref{fig:comp}A). For the analysis we used the {\fontfamily{lmss}\selectfont signal.spectrogram} package from the Python library {\fontfamily{lmss}\selectfont scipy}. The STFT analysis for Patients 1-6  shows some changes in the time-frequency data during the ictal state, but generally lacks the precision to determine seizure onset and cessation (except for Patient 3, where the spectral information consistently increases its intensity between frequencies 1-10 during the ictal spate).

%Permutation Entropy
We also explored a robust non-linear statistical measure that quantifies the complexity of a dynamical system through time series analysis of observations: the \textit{Permutation Entropy} (PE). Introduced by Bandt and Pompe in 2002 \cite{Bandt2002}, this method aims to capture the ordering relationships between the values of a time series by extracting a probability distribution of the ordinal patterns and then computing the Shannon entropy. Similarly to the STFT approach, we analysed the PE for each channel within the recordings and then calculated the mean across all channels to obtain a single representative series for each patient. For our analysis, we used the  method {\fontfamily{lmss}\selectfont permutation\_entropy} from the Python library {\fontfamily{lmss}\selectfont ordpy} \cite{pessa2021ordpy}. Figure \ref{fig:comp}B illustrates how PE measure shows high sensitivity to  seizures in MEG recordings (Patients 2 to 4), but it is not a reliable indicator for seizure detection  in iEEG and EEG recordings (Patients 1, 5 and 6).

%\begin{figure}[htb!]
%    \centering
%    \includegraphics[width=1\textwidth]{figures/New_figures/Figure_8.pdf}
    
%    \hspace{2pt}\includegraphics[width=8.05cm, height=1.95cm ]{figures/Spec_Beirut.png} \hspace{8.5pt}
%    \includegraphics[width=0.475\textwidth, ]{figures/PE_Beirut.png}
%    \caption{\textbf{Spectral and non-linear signal analysis.} Spectrograms associated to the Short-Time Fourier Transform (STFT) \textbf{(A)},  and averaged Permutation Entropy (PE) \textbf{(B)} from physiological recordings of Patients 1-6 (top to bottom). The ictal stages are indicated between the red bars and the red lines.}
%    \label{fig:comp}
%\end{figure}

\subsection{Local versus global seizures} 

Epileptic seizures may involve either part of the cerebral hemisphere (\textit{focal} seizures) or, on the contrary, the entire brain (\textit{generalized} seizures) --- the latter usually presenting loss of consciousness. This difference is reflected in either a local change in the brain dynamics, in which only a fraction of the set of the physiological signals presents a different behaviour, or a global change, in which all the signals present a synchronized seizure pattern. 

We used  Takens' theory  % to distinguish local and global seizures. Time-delay embeddings allow us
to derive a technique to topologically understand both focal and generalized seizures  (see \hyperref[takens]{Local vs global dynamics and delay-embeddings}). Concretely, we analyzed the \textit{global dynamic} induced by every individual signal (or observation) from a multichannel electrophysiological recording via its time-delay embedding. Next, we determined if they were consistent across the different channels or not. For generalized seizures, the global dynamic induced by every channel should coincide, whereas for focal seizures different local dynamics may arise.

\begin{figure}
    \centering
    \includegraphics[width=\textwidth]{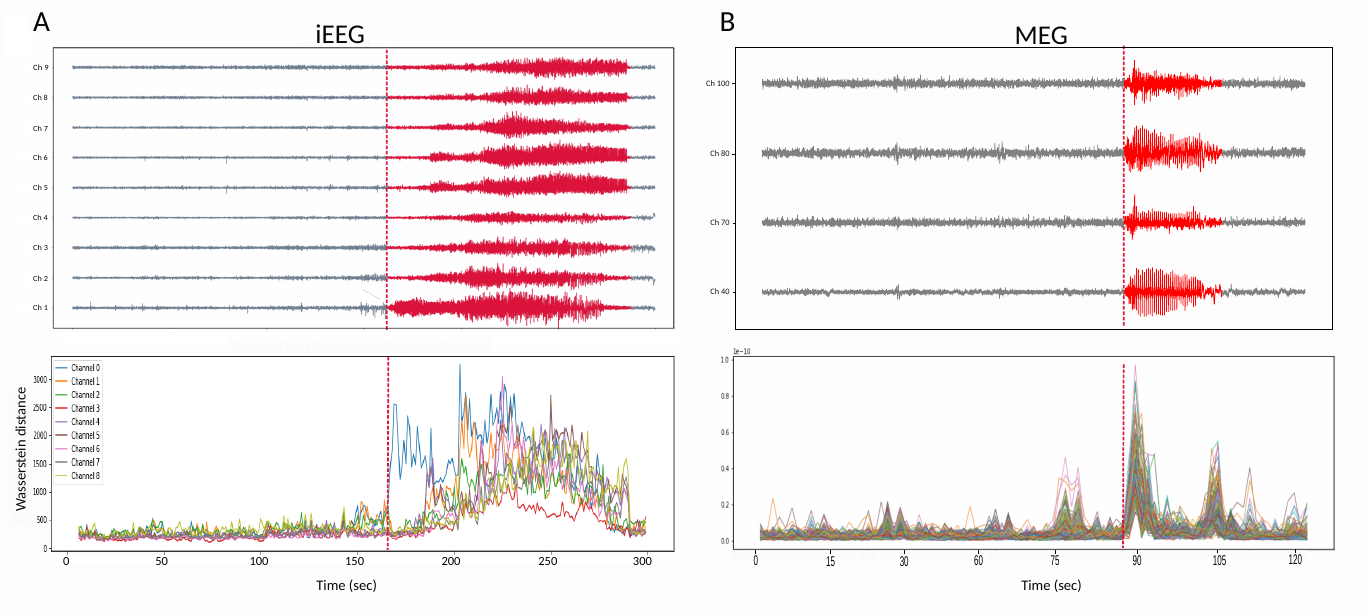}
    \caption{\textbf{The derivative biomarker for the persistence diagrams of the sliding-window time-delay embeddings.} \textbf{A.} Patient 1. There are salient peaks only in the channels associated to the focal seizure at the beginning ictal state (red dash line). \textbf{B.}\ Patient 3. There are synchronized peaks at the start  (red dash line) for all the channels. }
    \label{fig:takens}
\end{figure}

From a topological point of view, we were interested in the change of dynamics captured by the biomarker \eqref{derivative}. We computed the sliding-window delay embedding of recordings from Patients 1 and 3 (representing different types of seizures), with parameters $D=3$ and $\tau = 0.1$ seconds, for every channel\footnote{Here the choice of the parameters is made heuristically, although Taken's Theorem guarantees \textit{diffeomorphic} reconstructions for any $\tau>0$ and $D$ greater that twice the intrinsic dimension of the system (which is unknown). In practice, we tested a range of values for the parameters $D$ and $\tau$, all giving similar outputs}. Then, we overlapped the derivative functions \eqref{derivative} for every channel to compare the behavior (and changes) of the different induced dynamics (see Figure \ref{fig:takens}).

%\textcolor{red}{explain how to distinguish global vs local seizures with takens}

%\textcolor{red}{explain how to distinguish starting area}

%\textcolor{red}{explain the computations/details of parameters}

Patient 1 presents a focal seizure that generalizes. Figure \ref{fig:takens} (top) shows a first peak only at Channel 1, where the seizure starts, and then the derivative associated to rest of the channels starts to increase. In the same direction, the derivative of some channels, like Channel 5 and Channel 6, remains higher for a longer period of time given the persistence of the seizure dynamics on those signals. 
On the contrary, Patient 3 presents a generalized seizure. Figure \ref{fig:takens} (bottom) shows a consistent behavior of the all the channel during the entire ictal state.

\subsection{The size of the window}
The size of the window $w$ is a parameter that determines the length of the sub-interval of time in the multichannel recording to be embedded.
The choice of this parameter affects the level of refinement of the local properties from the dynamics that are captured in the embedding. Large values of $w$ provide coarse topological representations of the local dynamics, while small values offer finer granularity. There is a trade-off between the level of precision and the computational efficiency of the analysis (c.f. \hyperlink{computational implementation}{Computational Implementation}).

\begin{figure}[htb!]
\centering
 \includegraphics[width=\textwidth]{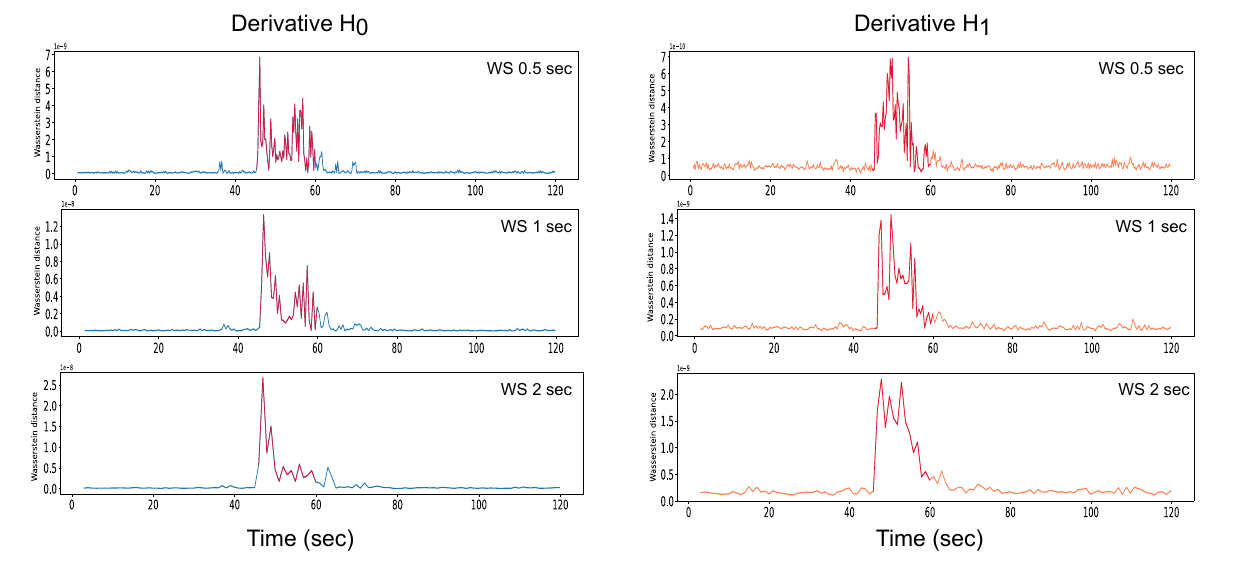}
\caption{\textbf{The \textit{window size} parameter.} Computation of the derivative biomarker for Patient 2 for $H_0$ (left) and $H_1$ (right) for different values of the window size parameter $w$.}
\label{fig:window_size}
\end{figure}

We explore the derivative biomarker \eqref{derivative} for different values of the window size $w$, keeping the overlap rate at $\frac{w}{2}$, for Patient 2 (Figure \ref{fig:window_size}). Notice that the size of the local embedding point clouds is $f \cdot w$ with $f$ being the frequency sample, and the actualization time is $\frac{w}{2}$. For $w=0.5$ seconds, the derivative shows almost instantaneous changes, with fast running time and actualization, while it is too sensitive to small changes. On the contrary, for $w=2$ seconds, the derivative shows averaged and smoothed changes, with slow running time and actualization. Our choice of $w=1$ seconds achieves balance between temporal resolution and computational efficiency.

%\begin{figure}[htb!]
%\centering
%\includesvg[width = 0.495 \textwidth]{figures/SW_RAO_consecutive_distance_H0_312.svg}    \includesvg[width = 0.495 \textwidth]{figures/SW_RAO_consecutive_distance_H1_312.svg}

%\includesvg[width = 0.495 \textwidth]{figures/SW_RAO_consecutive_distance_H0_625.svg}
%\includesvg[width = 0.495 \textwidth]{figures/SW_RAO_consecutive_distance_H1_625.svg}

%\includesvg[width = 0.495 \textwidth]{figures/SW_RAO_consecutive_distance_H0_1250.svg}%
%\includesvg[width = 0.495 \textwidth]{figures/SW_RAO_consecutive_distance_H1_1250.svg}
%    \caption{\textbf{The \textit{window size} parameter.} Computation of the derivative biomarker for Patient 2 for $H_0$ (left) and $H_1$ (right) for different values of the window size parameter $w$.}
%    \label{fig:window_size}
%\end{figure}

\subsection{Computational implementation} \label{computational implementation}

Our approach consists of the real time computation of the persistent diagrams associated to the time-evolving point clouds of approximate size of $\sim 500$ points (625 points for MEG recordings at sampling frequency 625 Hz, 400 points for iEEG recordings at frequency 200 Hz and  512 points for EEG recordings at frequency 256 Hz).
These diagrams are updated every 0.5 seconds. The execution time of the computation of every diagram is $\sim 0.25$ seconds and it is performed with the software {\fontfamily{lmss}\selectfont Ripser}  \cite{ripser2021}. All the computations were performed in a MacBook Pro 8-core CPU 32-core GPU 16 GB unified memory.

The 1-Wasserstein distance between persistence diagrams
can be computed using the Hungarian algorithm \cite{kuhn1955hungarian}, which is implemented in the package {\fontfamily{lmss}\selectfont Ripser} and in practice it takes $\sim 0.1$ seconds.

The total persistence of a persistence diagram can be computed in linear time in the number of points of a diagram, which in practice takes only $\sim 0.5\times 10^{-5}$ seconds.

Overall, our pipeline consists of the computation of the persistence diagram up to degree 1 of the sliding-window embedding of the signals, and the computation of the derivative and the total persistence indicators for every degree. All this process (which is updated  every 0.5 seconds)
takes less than 0.5 seconds
and it can be performed in real time.

%\cite{chen2021approximation}

%{\bf Data availability.} Codes used to simulate and fit the seizure dynamics are publicly available in GitHub.

\subsection{Statistical analysis  of the total persistence biomarker} \label{quantitative}
In this section, we conduct a statistical study to analyze the sensitivity of the Total Persistence biomarker  to epileptic signals. In contrast to classical algorithmic studies and global machine learning classifications, we  perform \textit{individualized} analysis for different patients. Indeed, the numerical topological summaries may vary from person to person, and in practice, the thresholds and parameters are adjusted individually for each patient. 
To complement our previous analysis in \href{top seizure detection}{Topological Seizure Detection} that aimed to capture changes  in the multidimensional dynamics from the recordings, we examine the statistical distribution of the values of the total persistence biomarker along the ictal and interictal states.
We build and analyze a datasets of disjoint segments  of 1-second EEG signals previously classified as either normal or epileptic. We use EEG recordings of brain activity of Patients 2, 5 and 6 from the databases from Toronto Western Hospital, CHB-MIT \cite{CHB-MIT} and Beirut Medical Center \cite{nasreddine2021epileptic}. Note that these patients are characterized by the weak effectiveness of standard signal analysis methods in detecting epileptic seizures from the EEG recording (see \href{comparison}{Comparison with standard seizure indicators}). Indeed, we compare the discriminate power of the Total Persistence biomarker with the Permutation Entropy signature.

\begin{figure}[htb!]
    \centering
    \includegraphics[width=\textwidth]{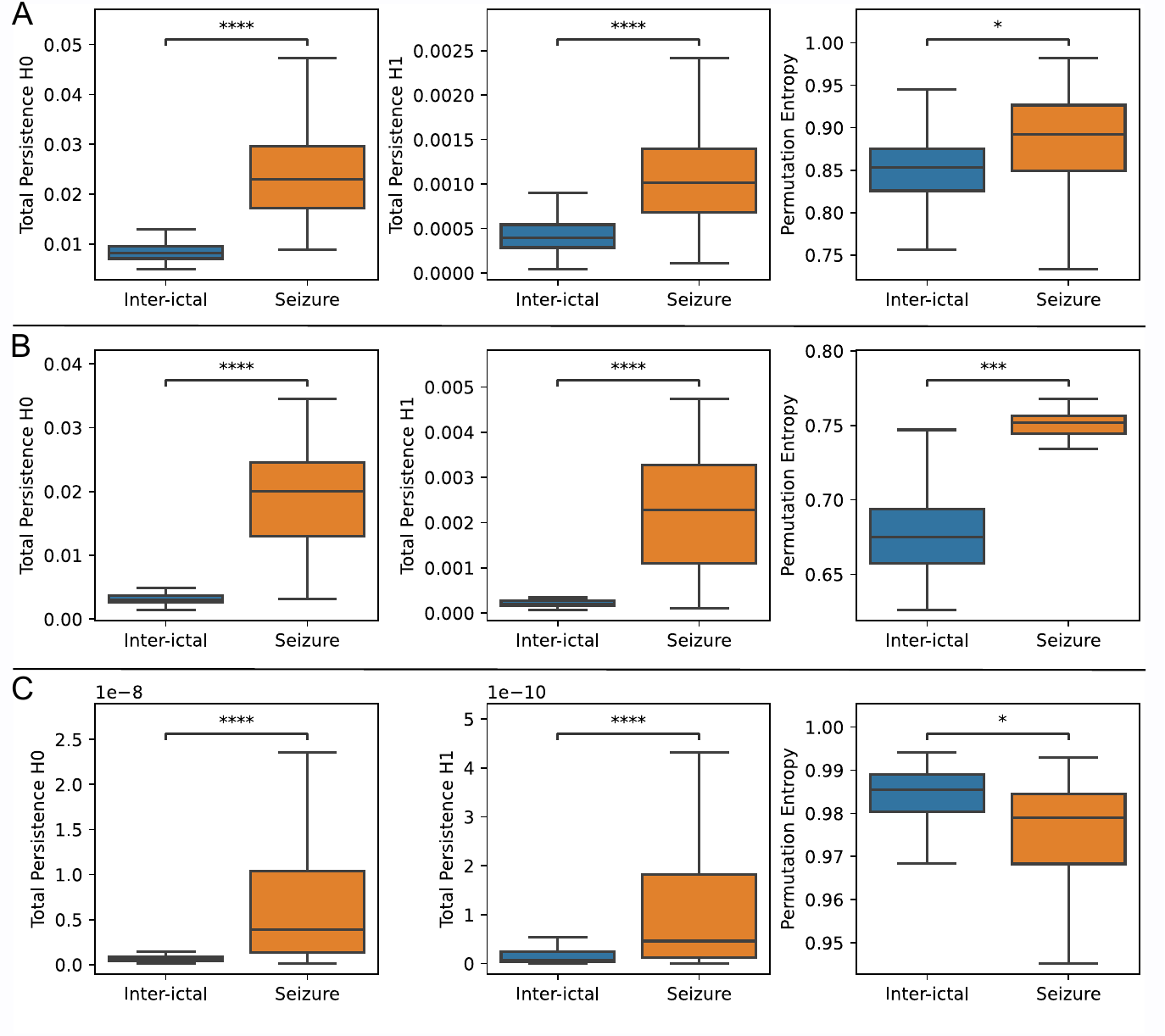}
    \caption{\textbf{The statistical significance of the  biomarkers.} The distribution of the total persistence summary for 1-second windows from EEG recordings from the ictal (seizure) and interictal state (baseline) in \textbf{A.} Patient 5, and \textbf{B.} Patient 6. \textbf{C.} Patient 2. Non-parametric statistical analysis via the Kruskal-Wallis test with Bonferroni correction was used to compare states, $NS~ not~ significant$, $*~p \leq 5.00e-02$, $** p \leq 1.00e-02$, $***~p \le 1.00e-03$, $**** ~p \le 1.00e-04$.}
    \label{fig:statistics}
\end{figure}

Simulating the continuous clinical monitoring process of individualized patients, we analyze databases of long recordings of the brain activity of Patient 2, including 7 seizures totaling 171 seconds, Patient 5, including 7 seizures totaling 442 seconds, and Patient 6, including 4 seizures totaling 111 seconds.
We preprocessed the databases to construct datasets consisting of an equal number \( N_{sec} \) of 1-second (non-overlapping) snippets of each inter-ictal and ictal state for each database (specifically, \( N_{sec}=171 \) for Patient 2, \( N_{sec}=442 \) for Patient 5, and \( N_{sec}=111 \) for Patient 6). For each 1-second signal segment, we calculate its persistence diagram and total persistence for $H_0$	and $H_1$, alongside the Permutation Entropy measure. Finally, non-parametric statistical analysis based on the Kruskal-Wallis method with Bonferroni correction \cite{kruskal1952use} is used to compare ictal vs inter-ictal states (see Figure \ref{fig:statistics}).

The results show that the Kruskal-Wallis test identifies a significant difference (with \( p \)-values lower than 0.0001) between the distribution values of the total persistence biomarkers during ictal vs. interictal states across the three patients analyzed. This highlights the robustness of the total persistence biomarkers in distinguishing between these states.  We emphasize that seizures in Patient 6 were only video-detected during the original data collection and analysis from \cite{nasreddine2021epileptic}, meaning that EEG visual inspection alone was not previously successful in identifying the seizures. 

In contrast, the Permutation Entropy signature distributions are weakly distinguishable (with \( p \)-values on the order of 0.05), except for Patient 6 where the \( p \)-value is lower than 0.001. In practice, due to the substantial overlap in the range of its distribution values, Permutation Entropy itself has low discriminative power.
These findings align with the inspection of change dynamics during particular seizures for Patients 5 and 6 (Figures \ref{fig:patient_CHB-MIT} and \ref{fig:patient_Beirut}), where the Permutation Entropy indicator also does not show prominent changes.

Overall, this experiment underscores the robustness of the total persistence biomarker to be integrated into an individualized system for real-time seizure tracking, given its consistent values for each individual.

\section{Discussion}

%We have presented a novel approach with roots in algebraic topology to detect epileptic seizures in real time from neurophysiological recordings. Most standard techniques, like time-frequency analysis or complexity measures, rely in the (averaged) properties of the set of time series. On the contrary, our indicators works with the dynamic information provided by the whole set of signals. Indeed, we identify the ictal state in terms of the changes in the geometry of the embeddings produced by the geometric change in the dynamics. The topological nature of our point of view provides consistency to different methods of collection of the neural information, such as EEG, iEEG and MEG, and to a range of clinical conditions.

Epilepsy is a neurological disorder characterized by recurrent seizures, which are sudden, uncontrollable electrical disturbances in the brain. It is a condition with a wide variety of manifestations, including focal, generalized, idiopathic, symptomatic, or cryptogenic, as well as subtypes and variants. Particularly, seizure detection using EEG/iEEG or MEG recordings faces numerous challenges. These range from distinguishing seizure activity from background noise or artefacts to the low signal-to-noise ratio in EEG/iEEG or MEG recordings, which can obscure subtle seizure activity, resulting in missed detections or false negatives. One of the most important problems is the variability of types of epilepsy, underlining the complexity of this neurological disorder and the importance of an individualized approach to its diagnosis and treatment. Therefore, we have proposed a method for the individual analysis of each patient, without making comparisons  between different clinical situations.

In this work, we  presented a novel methodology for analyzing electrophysiological signals from EEG/iEEG or MEG through the geometric study of multidimensional signal dynamics. Our approach leveraged  mathematical tools from algebraic topology, particularly persistent homology, to capture the topological features inherent in the time-varying dynamics of brain activity. By representing the signals as trajectories in high-dimensional spaces and examining  topological biomarkers, we gained insights into the underlying dynamics of epileptic seizures. This geometric perspective allowed us to identify distinctive interpretable patterns associated with pre-ictal, ictal, and post-ictal states, providing a deeper understanding of the seizure process. Moreover, our methodology offered the advantage of being applicable in real-time, enabling prompt detection and tracking of seizure events. Through extensive experimentation and validation on diverse datasets, including both scalp and intracranial recordings, we demonstrate the effectiveness and robustness of our approach across various clinical scenarios.

Most analysis methods are based on the study of one-dimensional signals, where each channel is analyzed separately and then averaged if a global understanding of the system is desired for ease of interpretation. However, as with any averaged information, there is an obvious loss of overall system information. Furthermore, focusing on the analysis of individual channels results in a loss of information about the underlying multidimensional dynamics in the brain. However, the lack of information about the relationships between channels can be addressed by connectivity (correlation) analysis. There are many methods for calculating connectivity, such as coherence, cross-correlation, mutual information, or phase locking. However, these methods suffer from the problem of losing the temporal information of the system, since they require averaging over the recording time. Our topological approach enables the study of temporal evolution in multidimensional dynamics, overcoming the mentioned limitations. Moreover, thanks to the persistent derivative biomarker, we can detect changes in these dynamics with great specificity, thus accurately identifying the beginning and end of seizures.

The algorithm presented in this work was compared with two of the most widely used methods in the literature: spectral decompositions of signals and Permutation Entropy signatures. Whereas those methods inconsistently identify sometimes the difference between ictal and interictal activities,  the proposed topological biomarkers demonstrated greater and consistent sensitivity in detecting seizures. This can be attributed to the ability of our algorithm to analyze the dynamics induced by all signals simultaneously, unlike conventional methods that rely on channel-wise averaging. However, there may be a need to understand the dynamics of individual channels separately, both to determine whether the epilepsy is partial or generalized and to try to identify the channel(s) from which the seizure originates. Based on this, our work also proposes a topological analysis of individual channels through the use of Takens' theorem, which allows the unambiguous detection of the channel where the seizure begins and subsequently generalizes.
Particularly remarkable is the case study of Patient 6, who had video-detected epilepsy but no displayed changes in the EEG. However, all topological biomarkers showed significantly higher values during the ictal state. This challenges the conventional notion that the ictal state of video-detected epilepsy cannot be detected on the EEG and suggests that topological analysis methods may provide valuable insights into the underlying signal dynamics.

We conducted individualized statistical analyses to compare (different types of) seizure signals  with control signals, fitted to each specific patient. %This approach could shed light on the complex dynamics of epileptic seizures, particularly regarding their manifestation on EEG recordings. 
The significant differences in the topological biomarkers observed between ictal and interictal data for every patient underscore the importance of utilizing advanced quantitative analysis techniques, such as persistence diagrams and associated biomarkers, to reveal the intricate dynamics of epileptic seizures, especially concerning their manifestation on EEG recordings. This enables the detection of subtle yet significant differences in signal dynamics. %Notably, in video-detected epilepsy, despite the absence of discernible visual changes in the EEG, the persistent topological derivative exhibited significantly higher values compared to other seizure types. 
%This challenges the conventional notion that the ictal state of video-detected epilepsy cannot be visually detected on the EEG and suggests that quantitative analysis methods may provide valuable insights into the underlying signal dynamics. 
%Furthermore, the differentiation between three seizure subtypes in terms of signal dynamics highlights the complexity of epileptic phenomena and the potential utility of quantitative EEG analysis in characterizing different seizure types.

While computational speed is a fundamental factor in developing a real-time epilepsy detection method, addressing the \textit{practical} implementation of the proposed algorithm in clinical settings is also essential. This entails discussing potential challenges and considerations for real-world deployment. Our method demonstrates robust computational performance for real-time updates, irrespective of the dimensionality of the problem (i.e., the number of electrodes). However, managing the dimensionality of recordings could present a challenge for channel-by-channel analysis. Sequential analysis algorithms may result in a linear increase in computation time with the addition of channels, or alternatively, parallel computation algorithms may be necessary. Therefore, while our algorithm does not necessitate parallel implementation from a technical standpoint, further investigation is required to evaluate its feasibility and scalability in clinical environments. This includes considerations such as hardware requirements, integration with existing EEG (ambulatory) monitoring systems, and validation in real patient scenarios.

%\textcolor{teal}{An important point to consider is the size of the window, as it cannot be too small to avoid losing relevant system information, nor too large to maintain specificity in dynamic changes. Nyquist's Theorem stipulates that in order to reconstruct an analog signal from its samples, it is necessary to sample the signal at a frequency at least twice as high as the highest frequency present in the original signal \cite{shannon1949communication}. However, some authors suggest that at least 20 points per cycle are needed to obtain good frequency information from the signal \cite{pikovsky2002synchronization}, while others argue that 7 points per cycle are sufficient. Based on the latter case, if we have a 1-second window with a sampling frequency of 200 Hz, we would have reliable signal information up to a frequency of approximately $(1*200)/7 \sim 30$ Hz. If we wanted to get information from higher frequencies, we would have to double or triple the window size. At higher sampling rates, the window for analysis can be smaller. However, significantly increasing the window size is not always optimal, as interictal-ictal dynamic changes occur over a very short period of time, leading to a loss of specificity in seizure detection with a very large window. This brings us back to the idea of personalizing the analysis for each patient, with the parameters of the analysis -- including the window size -- depending on the patient being treated and the equipment used to acquire the data.}
%  

The primary aim of this paper was to introduce a novel method of personalized analysis and to demonstrate its efficacy in different scenarios, including various recording types (EEG, iEEG, or MEG) and different epilepsy subclasses (focal or generalized, and even video detected). Despite the relatively small number of patients analyzed, our focus was on individualized analysis of the data for each patient through clear and interpretable biomarkers, rather than relying on a global automatic machine learning algorithm. However, a thorough analysis using a large clinical dataset for every patient is necessary to extrapolate to clinical practice.
Moreover, the next phase of this research will involve integrating machine learning techniques to fit the parameters of our procedure (e.g., size of the window, seizure-threshold for the biomarkers) to every patient, optimizing the accuracy of seizure detection. %By leveraging advanced computational methods, particularly deep learning, there is potential to enhance the sensitivity and specificity of seizure detection while also reducing computational time.

Overall, we believe that this novel method holds potential in medical practice as a valuable tool for clinical diagnosis. It promises enhanced precision in detecting the onset of the ictal state, thereby offering improved accuracy in epilepsy diagnosis. Moreover, it has the potential to greatly assist in pinpointing epileptic foci, facilitating more targeted and effective treatment strategies. In summary, our work contributes to the advancement of computational individualized techniques for seizure detection and offers promising avenues for improving epilepsy diagnosis and treatment.

\section{Conclusions}
Epilepsy presents a complex array of manifestations, necessitating an individualized approach to diagnosis and treatment. Our study introduces a novel methodology rooted in algebraic topology for analyzing electrophysiological signals to detect epileptic seizures in real time, offering a deeper understanding of multidimensional signal dynamics in the brain. While our proposed algorithm exhibits superior sensitivity in seizure detection compared to conventional methods, challenges remain in practical implementation, particularly in real-world clinical settings. Looking ahead, the integration of machine learning techniques %, notably deep learning, 
holds promise for further improving seizure detection accuracy and efficiency. These future directions underscore our commitment to advancing epilepsy research and ultimately enhancing patient care.

\bigskip

{\bf Acknowledgements.}  We are grateful to Eugenio Borghini
for many useful discussions and suggestions during the preparation of this article. X. F. is a member of the Centre for Topological Data Analysis funded by the EPSRC grant EP/R018472/1. D. M. wants to thank for the financial support from MinCyT-FonCyT  PICT-2019, N° 01750 PMO BID.
For the  purpose of Open Access, the authors have applied a CC BY public copyright licence to any  Author Accepted Manuscript (AAM) version arising from this submission.

\bibliographystyle{plain}
\bibliography{biblio}

\begin{thebibliography}{10}

\bibitem{A96}
Henry D.~I. Abarbanel.
\newblock {\em Analysis of observed chaotic data}.
\newblock Institute for Nonlinear Science. Springer-Verlag, New {Y}ork, 1996.

\bibitem{AB93}
Henry D.~I. Abarbanel, Reggie Brown, John~J. Sidorowich, and Lev~Sh. Tsimring.
\newblock The analysis of observed chaotic data in physical systems.
\newblock {\em Rev. Modern Phys.}, 65(4):1331--1392, 1993.

\bibitem{Bandt2002}
C.~Bandt and B.~Pompe.
\newblock {Permutation entropy: a natural complexity measure for time series.}
\newblock {\em Physical Review Letters}, 88(17):174102, 2002.

\bibitem{ripser2021}
Ulrich Bauer.
\newblock Ripser: efficient computation of {V}ietoris-{R}ips persistence barcodes.
\newblock {\em J. Appl. Comput. Topol.}, 5(3):391--423, 2021.

\bibitem{BW83}
Joan~S. Birman and R.~F. Williams.
\newblock Knotted periodic orbits in dynamical systems. {I}. {L}orenz's equations.
\newblock {\em Topology}, 22(1):47--82, 1983.

\bibitem{cohen2005stability}
David Cohen-Steiner, Herbert Edelsbrunner, and John Harer.
\newblock Stability of persistence diagrams.
\newblock In {\em Proceedings of the twenty-first annual symposium on Computational geometry}, pages 263--271, 2005.

\bibitem{CEHM10}
David Cohen-Steiner, Herbert Edelsbrunner, John Harer, and Yuriy Mileyko.
\newblock Lipschitz functions have {$L_p$}-stable persistence.
\newblock {\em Found. Comput. Math.}, 10(2):127--139, 2010.

\bibitem{cole2021quantitative}
Alex Cole, Gregory~J. Loges, and Gary Shiu.
\newblock Quantitative and interpretable order parameters for phase transitions from persistent homology.
\newblock {\em Physical Review B}, 104(10):104426, 2021.

\bibitem{D05}
Luis~Garcia Dominguez, Richard~A Wennberg, William Gaetz, Douglas Cheyne, O~Carter Snead, and Jose Luis~Perez Velazquez.
\newblock Enhanced synchrony in epileptiform activity? local versus distant phase synchronization in generalized seizures.
\newblock {\em Journal of neuroscience}, 25(35):8077--8084, 2005.

\bibitem{EH10}
Herbert Edelsbrunner and John~L. Harer.
\newblock {\em Computational topology: An introduction}.
\newblock American Mathematical Society, Providence, RI, 2010.

\bibitem{ELZ02}
Herbert Edelsbrunner, David Letscher, and Afra Zomorodian.
\newblock Topological persistence and simplification.
\newblock {\em Discrete Comput. Geom.}, 28(4):511--533, 2002.
\newblock Discrete and computational geometry and graph drawing (Columbia, SC, 2001).

\bibitem{FBMG21}
Ximena Fernandez, Eugenio Borghini, Gabriel Mindlin, and Pablo Groisman.
\newblock Intrinsic persistent homology via density-based metric learning, 2020.

\bibitem{flanary2023reliability}
James Flanary, Samuel~R Daly, Caitlin Bakker, Alexander~B Herman, Michael~C Park, Robert McGovern, Thaddeus Walczak, Thomas Henry, Theoden~I Netoff, and David~P Darrow.
\newblock Reliability of visual review of intracranial electroencephalogram in identifying the seizure onset zone: A systematic review and implications for the accuracy of automated methods.
\newblock {\em Epilepsia}, 64(1):6--16, 2023.

\bibitem{Ghrist08}
Robert Ghrist.
\newblock Barcodes: the persistent topology of data.
\newblock {\em Bull. Amer. Math. Soc. (N.S.)}, 45(1):61--75, 2008.

\bibitem{GL02}
Robert Gilmore and Marc Lefranc.
\newblock {\em The topology of chaos}.
\newblock Wiley-Interscience [John Wiley \& Sons], New {Y}ork, 2002.
\newblock Alice in Stretch and Squeezeland.

\bibitem{CHB-MIT}
Ary Goldberger, Luís Amaral, L.~Glass, Shlomo Havlin, J.~Hausdorg, Plamen Ivanov, R.~Mark, J.~Mietus, G.~Moody, Chung-Kang Peng, H.~Stanley, and Physiotoolkit Physiobank.
\newblock Components of a new research resource for complex physiologic signals.
\newblock {\em PhysioNet}, 101, 01 2000.

\bibitem{gotman1990automatic}
Jean Gotman.
\newblock Automatic seizure detection: improvements and evaluation.
\newblock {\em Electroencephalography and clinical Neurophysiology}, 76(4):317--324, 1990.

\bibitem{SohanianHaghighi2017}
Hossein~Sohanian Haghighi and Amir Hossein~Davaie Markazi.
\newblock A new description of epileptic seizures based on dynamic analysis of a thalamocortical model.
\newblock {\em Scientific Reports}, 7, 2017.

\bibitem{harpale2016time}
Varsha~K Harpale and Vinayak~K Bairagi.
\newblock Time and frequency domain analysis of {EEG} signals for seizure detection: A review.
\newblock In {\em 2016 International Conference on Microelectronics, Computing and Communications (MicroCom)}, pages 1--6. IEEE, 2016.

\bibitem{Hatcher02}
Allen Hatcher.
\newblock {\em Algebraic topology}.
\newblock Cambridge University Press, Cambridge, 2002.

\bibitem{hodgkin1952quantitative}
Alan~L Hodgkin and Andrew~F Huxley.
\newblock A quantitative description of membrane current and its application to conduction and excitation in nerve.
\newblock {\em The Journal of physiology}, 117(4):500, 1952.

\bibitem{Jirsa14}
Viktor~K. Jirsa, William~C. Stacey, Pascale~P. Quilichini, Anton~I. Ivanov, and Christophe Bernard.
\newblock {On the nature of seizure dynamics}.
\newblock {\em Brain}, 137(8):2210--2230, 06 2014.

\bibitem{karoly2017circadian}
Philippa~J Karoly, Hoameng Ung, David~B Grayden, Levin Kuhlmann, Kent Leyde, Mark~J Cook, and Dean~R Freestone.
\newblock The circadian profile of epilepsy improves seizure forecasting.
\newblock {\em Brain}, 140(8):2169--2182, 2017.

\bibitem{khoa2012detecting}
Truong Quang~Dang Khoa, Nguyen Thi Minh~Huong, Vo~Van Toi, et~al.
\newblock Detecting epileptic seizure from scalp eeg using lyapunov spectrum.
\newblock {\em Computational and Mathematical Methods in Medicine}, 2012, 2012.

\bibitem{kruskal1952use}
William~H Kruskal and W~Allen Wallis.
\newblock Use of ranks in one-criterion variance analysis.
\newblock {\em Journal of the American statistical Association}, 47(260):583--621, 1952.

\bibitem{kuhn1955hungarian}
Harold~W Kuhn.
\newblock The {H}ungarian method for the assignment problem.
\newblock {\em Naval research logistics quarterly}, 2(1-2):83--97, 1955.

\bibitem{lee2017quantifying}
Yongjin Lee, Senja~D Barthel, Pawe{\l} D{\l}otko, S~Mohamad Moosavi, Kathryn Hess, and Berend Smit.
\newblock Quantifying similarity of pore-geometry in nanoporous materials.
\newblock {\em Nature communications}, 8(1):1--8, 2017.

\bibitem{lehnertz2003seizure}
Klaus Lehnertz, Florian Mormann, Thomas Kreuz, Ralph~G Andrzejak, Christoph Rieke, Peter David, and Christian~E Elger.
\newblock Seizure prediction by nonlinear {EEG} analysis.
\newblock {\em IEEE Engineering in Medicine and Biology Magazine}, 22(1):57--63, 2003.

\bibitem{M16}
Emanuela Merelli, Marco Piangerelli, Matteo Rucco, and Daniele Toller.
\newblock A topological approach for multivariate time series characterization: the epileptic brain.
\newblock In {\em Proceedings of the 9th EAI international conference on bio-inspired information and communications technologies (formerly BIONETICS)}, pages 201--204, 2016.

\bibitem{mustafa2011analysis}
Mahfuzah Mustafa, Mohd~Nasir Taib, Zunairah~Hj Murat, Norizam Sulaiman, and Siti Armiza~Mohd Aris.
\newblock The analysis of {EEG} spectrogram image for brainwave balancing application using {ANN}.
\newblock In {\em 2011 UkSim 13th International Conference on Computer Modelling and Simulation}, pages 64--68. IEEE, 2011.

\bibitem{nasreddine2021epileptic}
W~Nasreddine.
\newblock Epileptic {EEG} dataset.
\newblock {\em Mendeley Data V1}, 1, 2021.

\bibitem{nicolaou2012detection}
Nicoletta Nicolaou and Julius Georgiou.
\newblock Detection of epileptic electroencephalogram based on permutation entropy and support vector machines.
\newblock {\em Expert Systems with Applications}, 39(1):202--209, 2012.

\bibitem{nicolau2011topology}
Monica Nicolau, Arnold~J Levine, and Gunnar Carlsson.
\newblock Topology based data analysis identifies a subgroup of breast cancers with a unique mutational profile and excellent survival.
\newblock {\em Proceedings of the National Academy of Sciences}, 108(17):7265--7270, 2011.

\bibitem{P18}
Yash Paul.
\newblock Various epileptic seizure detection techniques using biomedical signals: a review.
\newblock {\em Brain informatics}, 5(2):1--19, 2018.

\bibitem{P19}
Jose~A. Perea.
\newblock Topological times series analysis.
\newblock {\em Notices Amer. Math. Soc.}, 66(5):686--694, 2019.

\bibitem{PH15}
Jose~A. Perea and John Harer.
\newblock Sliding windows and persistence: an application of topological methods to signal analysis.
\newblock {\em Found. Comput. Math.}, 15(3):799--838, 2015.

\bibitem{PDG13}
Piero Perucca, François Dubeau, and Jean Gotman.
\newblock {Intracranial electroencephalographic seizure-onset patterns: effect of underlying pathology}.
\newblock {\em Brain}, 137(1):183--196, 10 2013.

\bibitem{pessa2021ordpy}
Arthur A.~B. Pessa and Haroldo~V. Ribeiro.
\newblock ordpy: A {P}ython package for data analysis with permutation entropy and ordinal network methods.
\newblock {\em Chaos: An Interdisciplinary Journal of Nonlinear Science}, 31(6):063110, 2021.

\bibitem{PRTM18}
Marco Piangerelli, Matteo Rucco, Luca Tesei, and Emanuela Merelli.
\newblock Topological classifier for detecting the emergence of epileptic seizures.
\newblock {\em BMC research notes}, 11(1):1--7, 2018.

\bibitem{rabadan2019topological}
Ra{\'u}l Rabad{\'a}n and Andrew~J. Blumberg.
\newblock {\em Topological data analysis for genomics and evolution: topology in biology}.
\newblock Cambridge University Press, Cambridge, 2019.

\bibitem{radhakrishnan1998estimating}
N.~Radhakrishnan and B.N. Gangadhar.
\newblock Estimating regularity in epileptic seizure time-series data.
\newblock {\em IEEE engineering in medicine and biology magazine}, 17(3):89--94, 1998.

\bibitem{sackellares2000epilepsy}
J~Chris Sackellares, Leontdas~D Iasemidis, Deng-Shan Shiau, Robin~L Gilmore, and Steven~N Roper.
\newblock Epilepsy--when chaos fails.
\newblock In {\em Chaos in Brain?}, pages 112--133. World Scientific, 2000.

\bibitem{shorvon2012oxford}
Simon Shorvon, Renzo Guerrini, Mark Cook, and Samden Lhatoo.
\newblock {\em Oxford textbook of epilepsy and epileptic seizures}.
\newblock OUP, Oxford, 2012.

\bibitem{ST20}
Primoz Skraba and Katharine Turner.
\newblock Wasserstein stability for persistence diagrams, 2020.

\bibitem{S67}
Stephen Smale.
\newblock Differentiable dynamical systems.
\newblock {\em Bulletin of the American mathematical Society}, 73(6):747--817, 1967.

\bibitem{Tak81}
Floris Takens.
\newblock Detecting strange attractors in turbulence.
\newblock In {\em Dynamical systems and turbulence, {W}arwick 1980 ({C}oventry, 1979/1980)}, volume 898 of {\em Lecture Notes in Math.}, pages 366--381. Springer, Berlin-New York, 1981.

\bibitem{veisi2007fast}
Iman Veisi, Naser Pariz, and Ali Karimpour.
\newblock Fast and robust detection of epilepsy in noisy eeg signals using permutation entropy.
\newblock In {\em 2007 IEEE 7th international symposium on bioinformatics and bioengineering}, pages 200--203. IEEE, 2007.

\bibitem{V11}
Jos{\'e} Luis~Perez Velazquez, Luis~Garcia Dominguez, Vera Nenadovic, and Richard~A Wennberg.
\newblock Experimental observation of increased fluctuations in an order parameter before epochs of extended brain synchronization.
\newblock {\em Journal of biological physics}, 37(1):141--152, 2011.

\bibitem{whitney1936differentiable}
Hassler Whitney.
\newblock Differentiable manifolds.
\newblock {\em Annals of Mathematics}, pages 645--680, 1936.

\bibitem{W74}
Robert~F. Williams.
\newblock Expanding attractors.
\newblock {\em Publications Math{\'e}matiques de l'IH{\'E}S}, 43:169--203, 1974.

\bibitem{WHO22}
{World Health Organization}.
\newblock Fact sheet on epilepsy.
\newblock {\em Online}, 2022.

\bibitem{yao2009topological}
Yuan Yao, Jian Sun, Xuhui Huang, Gregory~R. Bowman, Gurjeet Singh, Michael Lesnick, Leonidas~J. Guibas, Vijay~S. Pande, and Gunnar Carlsson.
\newblock Topological methods for exploring low-density states in biomolecular folding pathways.
\newblock {\em The Journal of chemical physics}, 130(14):04B614, 2009.

\bibitem{zabidi2012short}
A~Zabidi, W~Mansor, YK~Lee, and CWNF Che~Wan Fadzal.
\newblock Short-time {F}ourier {T}ransform analysis of {EEG} signal generated during imagined writing.
\newblock In {\em 2012 International Conference on System Engineering and Technology (ICSET)}, pages 1--4. IEEE, 2012.

\bibitem{zomorodian_2005}
Afra Zomorodian and Gunnar Carlsson.
\newblock Computing persistent homology.
\newblock {\em Discrete and Computational Geometry}, 33:249--274, 02 2005.

\end{thebibliography}

\iffalse
\section*{Author contributions statement}
 X.F. contributed to the design of the work, the acquisition of data, the investigation and statistical interpretation of results, creation of software, visualization, and writing the original draft. D. M. contributed to the design of the work, to the acquisition and processing of data, the conception of the project and the contextualization of the work, and the review of the manuscript.
 \fi

%\section{Data availability statement}

%\section{Additional information}
%The authors declare that they have no competing interests.

\end{document}